\documentclass[twocolumn,twocolappendix]{aastex6}

\usepackage{graphicx}
\usepackage{rotating}
\newcommand\teff{$T_{\mathrm{eff}}$}
\newcommand\logg{$\log g$}
\newcommand\vsini{$v \sin{i}$}

\newcommand\rp{$R_{\mathrm{p}}$}
\newcommand\rearth{$R_{\oplus}$}
\newcommand\sigmaew{$\sigma_{\mathrm{EW}}$}
\newcommand\sigmaeweq{\sigma_{\mathrm{EW}}}

\begin{document}

\title{Identifying Young $Kepler$ Planet Host Stars from Keck-HIRES Spectra of Lithium\altaffilmark{1}}
\author{Travis A.\ Berger\altaffilmark{2}, Andrew W.\ Howard\altaffilmark{2,3}, \& Ann Merchant Boesgaard\altaffilmark{2}}
\altaffiltext{2}{Institute for Astronomy, University of Hawaii, 2680 Woodlawn Drive, Honolulu, Hawaii 96822, USA}
\altaffiltext{3}{Department of Astrophysics, California Institute of Technology, MC 249-17, Pasadena, California 91125, USA}

\begin{abstract}
The lithium doublet at 6708 \AA\ provides an age diagnostic for main sequence FGK dwarfs. We measured the abundance of lithium in 1305 stars with detected transiting planets from the Kepler Mission using high-resolution spectroscopy. Our catalog of lithium measurements from this sample have a range of abundance from A(Li) = 3.11 $\pm$ 0.07 to an upper limit of $-$0.84 dex. For a magnitude-limited sample that comprises 960 of the 1305 stars, our Keck-HIRES spectra have a median S/N = 45 per pixel at $\sim$6700 \AA\ with spectral resolution $\frac{\lambda}{\Delta \lambda}$ = $R$ = 55,000. We identify 80 young stars that have A(Li) values greater than the Hyades at their respective effective temperatures; these stars are younger than $\sim$650 Myr old, the approximate age of the Hyades. We then compare the distribution of A(Li) with planet size, multiplicity, orbital period, and insolation flux. We find larger planets preferentially in younger systems, with an A-D two-sided test p-value = 0.002, a $>3\sigma$ confidence that the older and younger planet samples do not come from the same parent distribution. This is consistent with planet inflation/photoevaporation at early ages. The other planet parameters ($Kepler$ planet multiplicity, orbital period, and insolation flux) are uncorrelated with age.
\end{abstract}


\section{Introduction}
\footnotetext{Based on observations obtained at the W.\,M.\,Keck Observatory, 
                      which is operated jointly by the University of California and the 
                      California Institute of Technology.  Keck time has been granted by  
                      the University of Hawaii, the University of California, and Caltech.}
NASA's $Kepler$ Mission was designed to detect transiting planets and to measure the fraction of Sun-like stars with Earth-sized planets in the habitable zone. During the four year mission, $Kepler$ discovered more than 4000 exoplanet candidates, of which 2327 have been confirmed \citep{Coughlin2016}. Twenty-one of these confirmed exoplanets are 1--2$\times$ Earth size and orbit in the traditionally defined habitable zone. Analysis of $Kepler$ data demonstrated that 50\% of Sun-like stars harbor a planet between the size of Earth and Neptune with orbital periods less than 85 days \citep{Fressin2013}. Complementary Doppler surveys of nearby stars showed that 8.5\% of giant planets with periods shorter than a few years orbit similar type stars \citep{Cumming2008}. Studies by \cite{Howard2010}, \cite{Mayor2011}, and \cite{Howard2012} have shown that giant planets are less plentiful than their smaller counterparts. In addition, $Kepler$ analysis uncovered a diverse set of exoplanetary systems, some of which have peculiar properties and architectures. Noteworthy systems include the two habitable zone planets orbiting $Kepler$-62 \citep{Borucki2013}, the Earth-size planet with an 8.5 hour period orbiting $Kepler$-78 \citep{Sanchis-Ojeda2013}, and the $Kepler$-47 circumbinary system \citep{Orosz2012}. In each of these cases, measuring the stellar properties (e.g., radii, masses, and effective temperatures) is critical to determine the planet properties. For instance, all transit-derived planet radii scale directly with the stellar radius. Among the stellar properties, age is frequently unknown or poorly determined. Age is important because dynamic processes including mass loss, contraction, reinflation, and migration sculpt the planet population that we observe today.

Accurately determining stellar ages is difficult. The precise age of 4.567 Gyr for the sun is based on isotopic measurements of meteorites \citep{Chaussidon2007}, a method that is unavailable for other stars. \cite{Soderblom2010} provides a comprehensive review of the techniques to determine approximate stellar ages, including (1) kinematics, (2) isochrone placement through measured temperature, metallicity, and luminosity, (3) asteroseismology, (4) rotation rate, (5) magnetic activity, (6) lithium abundance, and (7) nucleocosmochronometry.

Some of these methods are based on only a few assumptions, but are observationally demanding. For example, nucleocosmochronometry requires high-resolution, high-signal-to-noise spectra and kinematic techniques need large groups of stars. Isochrone placement using precise temperatures, metallicity, and luminosity (together with their uncertainties) and asteroseismology are model-dependent methods that rely on detailed stellar physics. Sometimes, even with high quality observational data, astronomers cannot determine an isochrone age for stars using either method, due to poor interpolation between models and unresolved degeneracies in the Hertzprung-Russell Diagram. Empirical methods involving stellar rotation and magnetic activity are limited by calibration, measurement precision, and intrinsic astrophysical variability. 

Surface lithium abundance provides another age diagnostic. \cite{Herbig1965} was one of the first to consider Li as an age diagnostic for F and G stars. As Li is destroyed in the stellar interior at temperatures of $2.5 \times 10^6$ K primarily by ($p$,$\alpha$) reactions, surface Li abundance declines with time. The rate of decline is not uniform because transport mechanisms including convection and gravitational settling depend on effective temperature \citep{Xiong2009}. Lithium abundance can be measured by using the resonance doublet at 6708 \AA\ of Li I in stars. Measuring Li is observationally convenient because our high-resolution optical spectra used to determine bulk parameters (\teff, \logg, [Fe/H]) include the Li feature. Additionally, Li ages have been calibrated with measurements of several clusters. Unfortunately, precise ages are difficult to establish from sole analysis of the Li feature, but its presence is a discriminator of youth at least.

Others, such as \cite{Israelian2009}, \cite{Baumann2010}, \cite{Sousa2010}, \cite{Ramirez2012}, \cite{Mena2014,Mena2015}, \cite{Figueira2014}, and \cite{Gonzalez2014,Gonzalez2015}, have compared stellar Li abundance for exoplanet hosts versus single stars. \cite{Israelian2009} studied a uniform sample of 451 stars in the HARPS high precision radial velocity survey, with stars spanning \teff\ = 4900--6500 K. The authors found low Li abundance (A(Li) $\equiv$ 12 + $\log(\mathrm{Li/H})$) for stars in a narrow temperature range (\teff\ = 5700--5850 K) compared to stars without exoplanet companions, while excluding metallicity, age, \vsini, and activity as possible causes for this anomaly. They hypothesized mechanisms to account for this trend:  stars with planets might experience a) a different evolution, b) planets might infall and cause stellar mixing, and c) there may be interaction during the pre-main sequence (PMS) phase which can force high differential rotation and therefore enhanced Li depletion within planet-host stars.

However, some more recent work \citep{Baumann2010,Ramirez2012} contradicted the results of \cite{Israelian2009}, while others \citep{Sousa2010,Mena2014,Mena2015,Figueira2014,Gonzalez2014,Gonzalez2015} found supporting evidence for enhanced host star Li depletion. \cite{Baumann2010} studied a sample of 117 solar-type stars, 14 of which were planet-hosts. These stars exhibited normal A(Li) for their ages. In addition, the authors showed that 82 stars originally reported in the literature to support enhanced Li depletion in fact had normal A(Li) for their ages. \cite{Baumann2010} provide a few reasons for the disagreement between their results and \cite{Israelian2009}:  (1) the HARPS sample of solar analogs at [Fe/H] $\simeq$ 0.0 are on average older than non-planet-host stars, (2) metal-rich solar analogs are more lithium-poor than solar metallicity stars, and (3) the sample includes a number of peculiarly high Li abundances.

\cite{Ramirez2012}, like \cite{Baumann2010}, found that any connection between Li abundance and planet occurrence is likely a product of sample bias in stellar mass, age, and metallicity. \cite{Ramirez2012} studied a sample of 1381 dwarf and subgiant stars, 165 of which were planet-hosts. The large sample size allowed them to analyze trends in A(Li) with the presence of exoplanets, but the planet hosts and non-hosts were taken from different sources, and therefore could suffer from inhomogeneities. Their data suggest there is some planet-star interaction (not necessarily planet formation-related) that prevents planet-host stars from experiencing the sudden drop in A(Li) responsible for the Li desert, a region in A(Li)-\teff\ space where stars should appear empirically, but do not. Ultimately, \cite{Ramirez2012} rejected the presence of enhanced Li depletion in planet-hosts proposed by \cite{Israelian2009} after claiming to properly account for all possible sources of bias.

Unlike \cite{Baumann2010} and \cite{Ramirez2012}, a number of other studies continued to find enhanced Li depletion in host stars \citep{Sousa2010,Mena2014,Mena2015,Figueira2014,Gonzalez2014,Gonzalez2015}. \cite{Sousa2010} investigated potential effects of age and mass on Li depletion  and found that differences in ages and stellar mass could not explain the Li deficit in planet host stars. \cite{Gonzalez2014,Gonzalez2015} introduced new high resolution spectra of late-F and early-G stars, determined A(Li), and then added homogeneous literature data, finding that Li is deficient in giant planet hosts compared to comparison stars.

\cite{Mena2014} focused again on solar-type stars, finding that solar twins with hot jupiters show enhanced Li depletion compared to those without planets. In contrast to \cite{Ramirez2012}, \cite{Mena2014} utilized a homogeneous sample, entirely from HARPS and including both stars with planets and those without, to minimize potential confounding effects in A(Li). \cite{Figueira2014} used multivariable regression to test these confounding effects on previously published A(Li), and found that, when one assumes linearity in the fundamental stellar parameters, an offset in A(Li) between hosts and non-hosts is recovered. This offset is strongly statistically significant, but it is reduced to zero if host stars are replaced with comparison stars.

Finally, \cite{Mena2015} found a similar trend of Li depletion in late-F stars (\teff\ = 5900--6300 K), although the differences in A(Li) between hosts and stars with no detected planets are smaller in magnitude than for solar-type stars. However, the authors found that hot jupiter hosts had a higher average \vsini\ than the comparison stars, so the enhanced Li depletion could be explained by rotationally-induced mixing and not the presence of planets. Given the studies following \cite{Ramirez2012} and the care taken to minimize contamination in the HARPS sample, the observational evidence for decreased Li in giant planet hosts is convincing.

No study has yet used A(Li) to differentiate between a large sample of young and old exoplanetary systems. Therefore, we present the first large scale (N $>$ 1000) study of Li in $Kepler$ planet host stars that separates the population into young and old age groups. This allows us to investigate planetary evolution and the dynamic processes (migration, mass loss, contraction, reinflation, etc.) that sculpt the observed planet population. Moreover, this analysis adds another impactful dimension to the current parameter space of exoplanets heavily characterized by mass, radius, and effective temperature.

In Section 2, we discuss the California-Kepler Survey sample. Section 3 details our pipeline to determine A(Li) and each of the important tasks performed therein, including normalization, Doppler shifting, measurement of the equivalent width of Li, and the A(Li) computation. Section 4 analyzes the full catalog and searches for any trends in exoplanetary parameters with age. In Section 5, we discuss our results and provide an astrophysical interpretation of our findings.

\begin{figure}[t]
\includegraphics[width=3.35in]{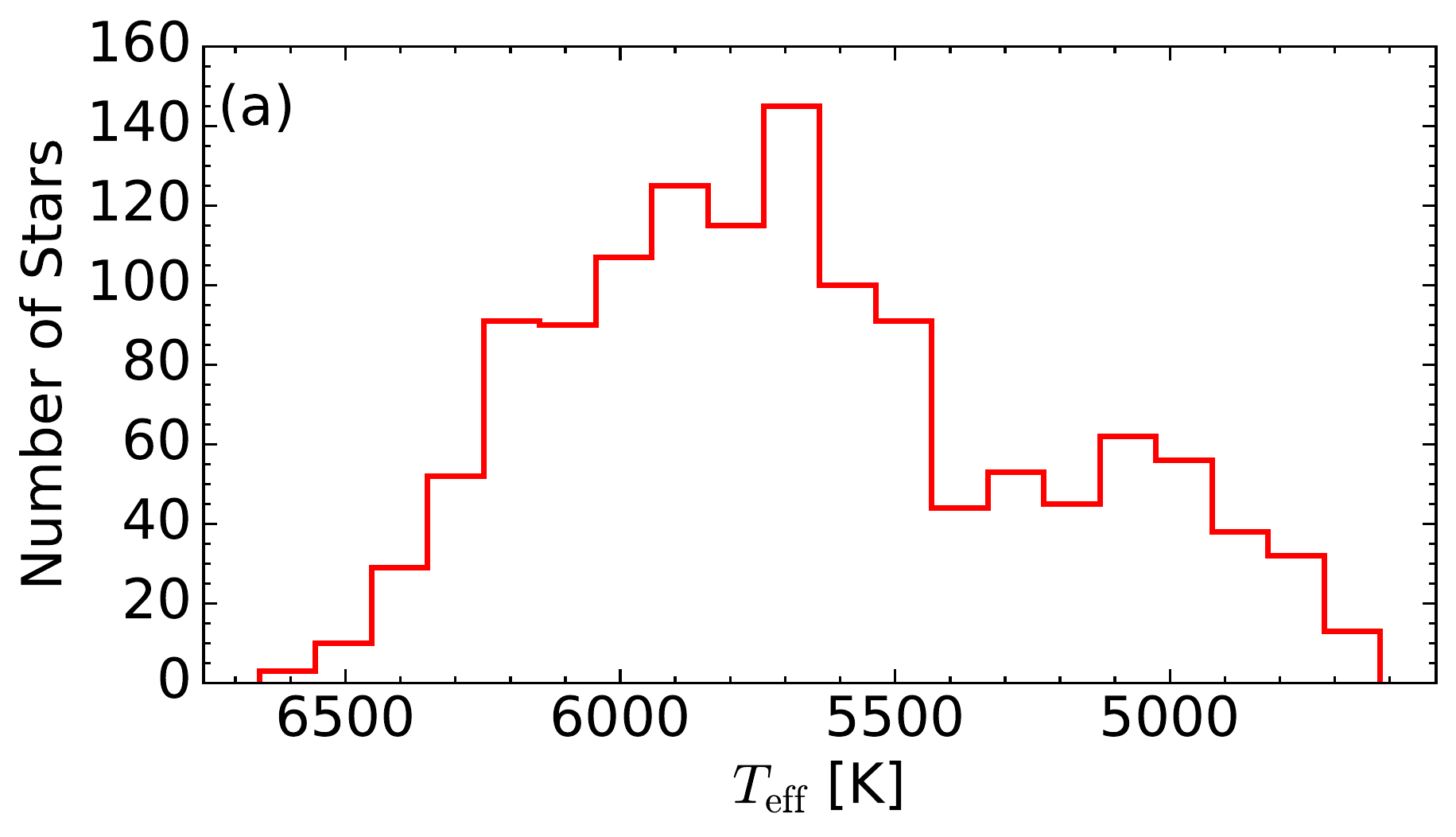}
\includegraphics[width=3.35in]{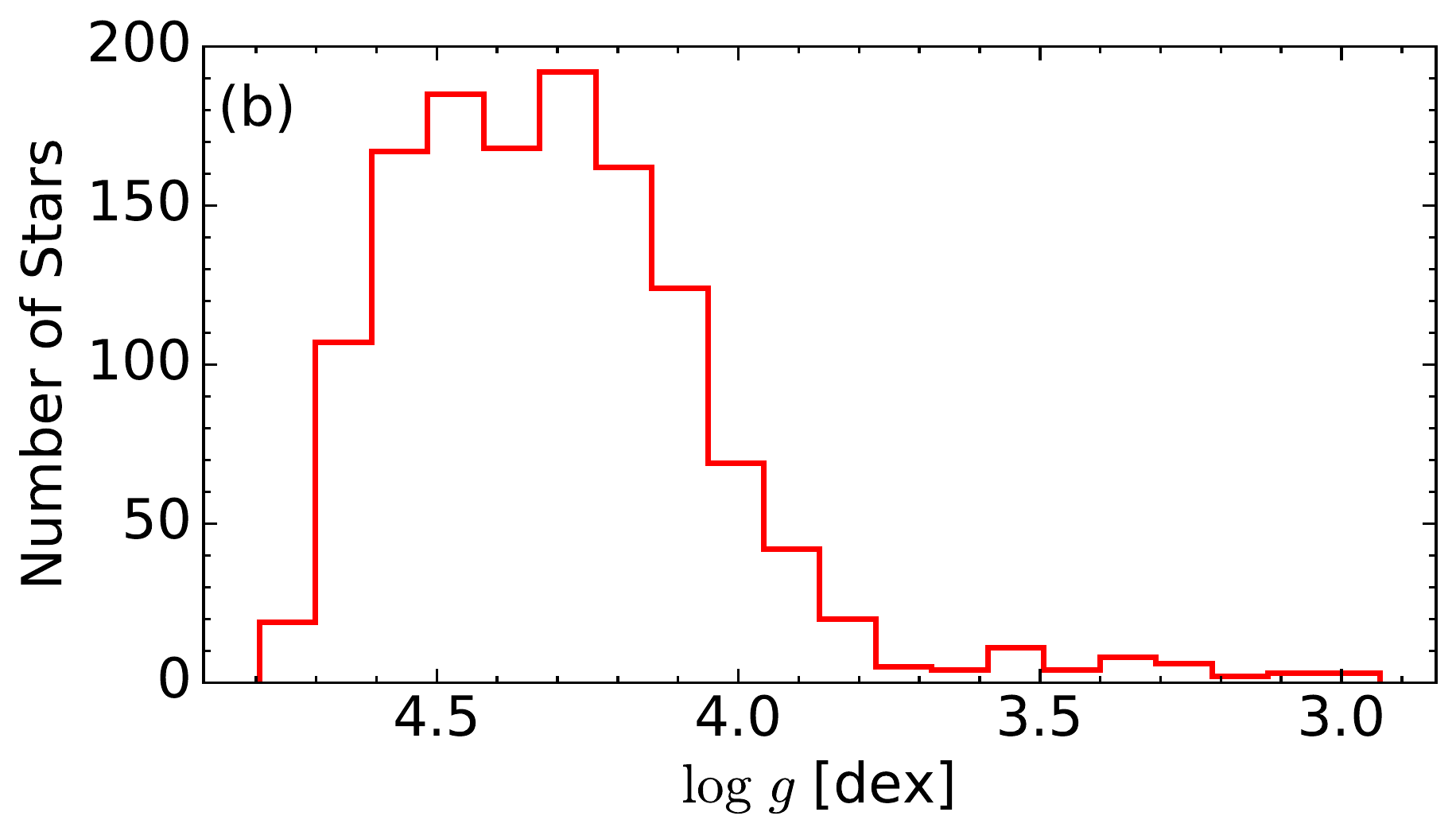}
\includegraphics[width=3.35in]{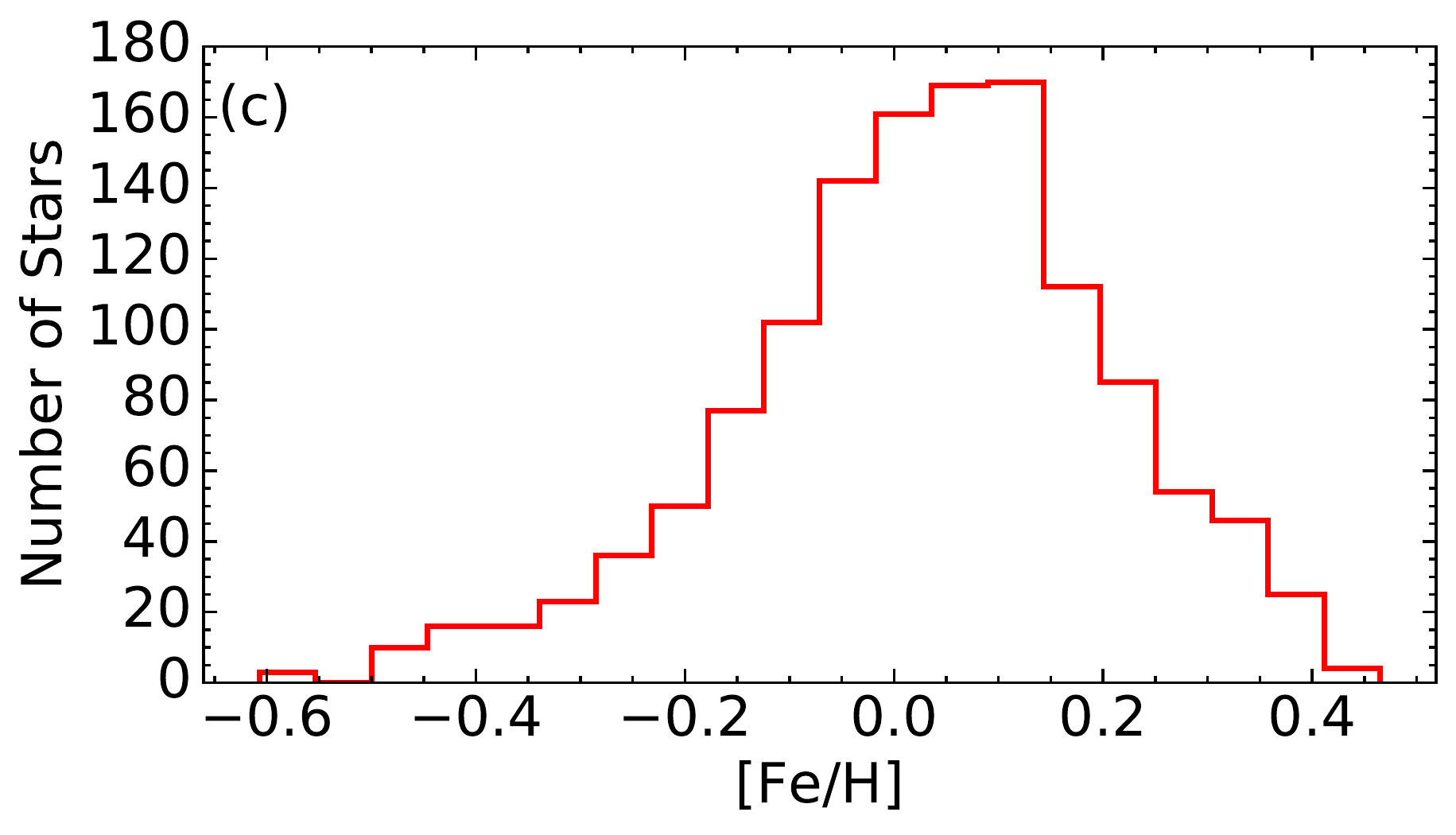}
\includegraphics[width=3.35in]{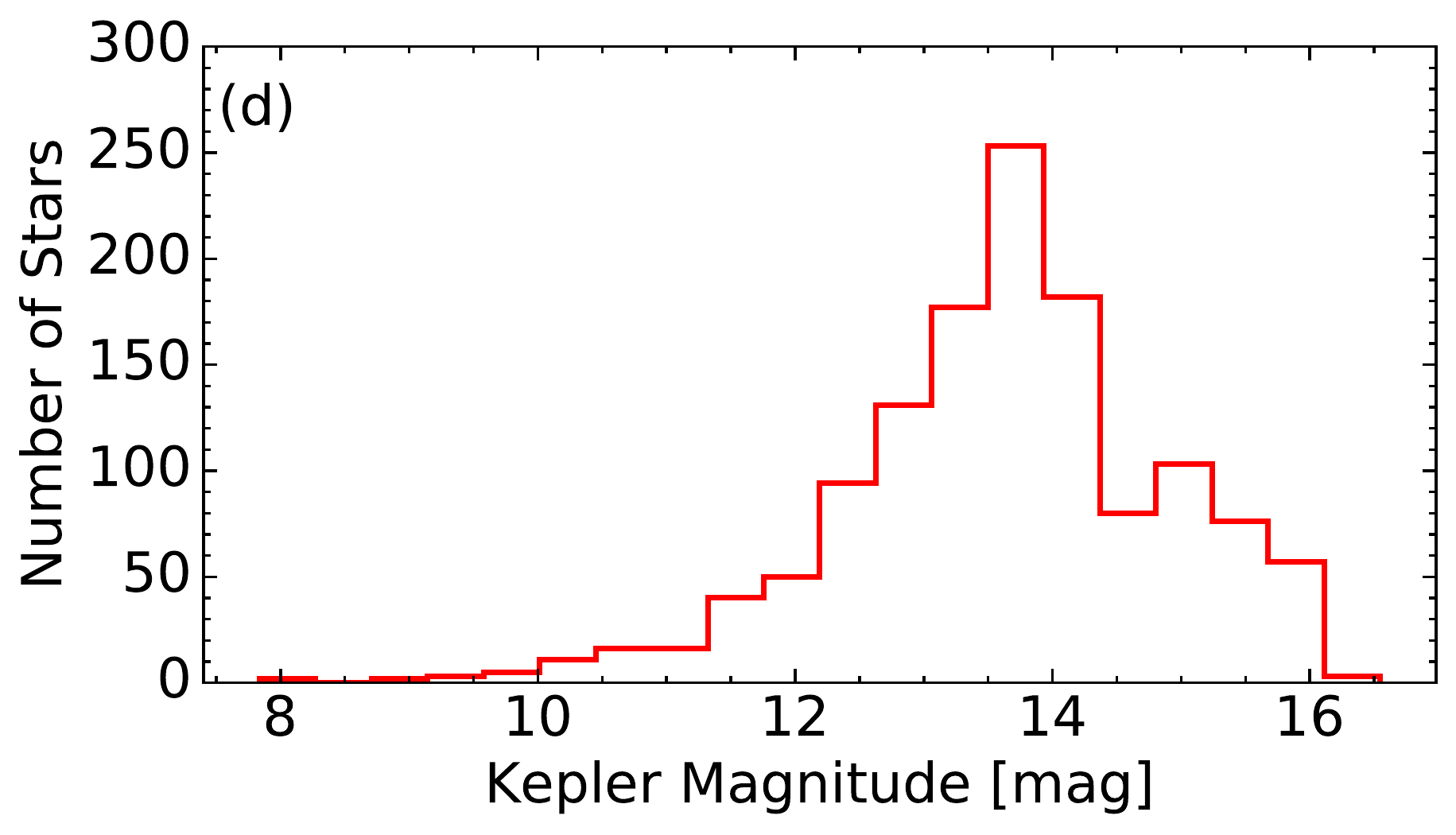}
\caption{Stellar sample. $(a)$: Histogram showing \teff\ for the 1305 CKS stars. $(b)$: Histogram of \logg\ for the CKS sample. $(c)$: [Fe/H]. $(d)$: Distribution of Kepler magnitudes. The majority of the stars have KepMag $\lesssim$ 14.23 mag.}\label{fig:CKS}
\end{figure}

\section{Stellar Sample}

One key follow-up survey of $Kepler$-discovered exoplanets is the California Kepler Survey (CKS) \citep{Petigura2017}, which was proposed to measure precise stellar parameters (\teff, \logg, [Fe/H], \vsini) by using local thermodynamic equilibrium (LTE) modeling of Keck-HIRES spectra of $\sim$1000 $Kepler$ FGK stars. Most of these stars are main sequence G and K dwarfs, but there are a few F stars. Figure \ref{fig:CKS} shows the distribution of our sample in \teff, \logg, and [Fe/H] histograms in plots (a), (b), and (c), respectively. The apparent magnitudes of the stars go down to 17th magnitude. Most spectra have signal-to-noise ratios ($S/N$) of $\sim$45 per pixel, or $\sim$90 per resolution element at 6700 \AA, with a resolution $R$ = 55,000 and wavelength coverage from 3642--7990 \AA. $S/N$ range from $\sim$5 to $\sim$200. We note that the primary CKS sample is magnitude-limited to a $Kepler$ magnitude (KepMag) $\lesssim$ 14.23 mag, with additional fainter stars from interesting groups, $i.e.$ habitable zone candidates and multi-planet systems. See Figure \ref{fig:CKS} plot (d) for the distribution of $Kepler$ magnitudes.

The spectra were reduced by removing cosmic rays, flat-fielding, bias subtraction, trimming, and column collapsing into a 1-D spectrum. We adopt the spectroscopic parameters (\teff, \logg, [Fe/H], \vsini) from \cite{Petigura2017}, computed from SpecMatch \citep{Petigura2015} and Spectroscopy Made Easy \citep{Valenti2012}. The spectral format of HIRES was kept fixed with 1--2 pixel accuracy for all spectra.

\section{Lithium Abundance Measurements}
We begin with the reduced HIRES spectra from \cite{Petigura2017} as detailed in the previous paragraph. To efficiently determine Li abundances for all stars within the CKS sample, we created an automated Li pipeline, which we detail below. The pipeline's spectrum analysis tasks include continuum normalization, Doppler correction, measurement of the Li equivalent width (EW), interpolation of a model atmosphere, determination of A(Li), and calculation of uncertainties ($\sigma_{\mathrm{A(Li)}}$).

\subsection{Continuum Normalization and Doppler Correction}
First, we utilized PyRAF's $continuum$ routine to remove the blaze function present in every spectrum. We applied this technique with the following options:  a 50-piece cubic spline fit, a low rejection criterion of 2.0$\sigma$, a high rejection criterion of 3.0$\sigma$, and 50 outlier rejection iterations. The output (normalized) spectrum is the input spectrum divided by the continuum-fit spline function. Outlier rejection allows $continuum$ to ignore any biasing effects from peaks (remaining cosmic rays) and troughs (absorption lines).

\begin{figure}[t]
\includegraphics[width=3.35in]{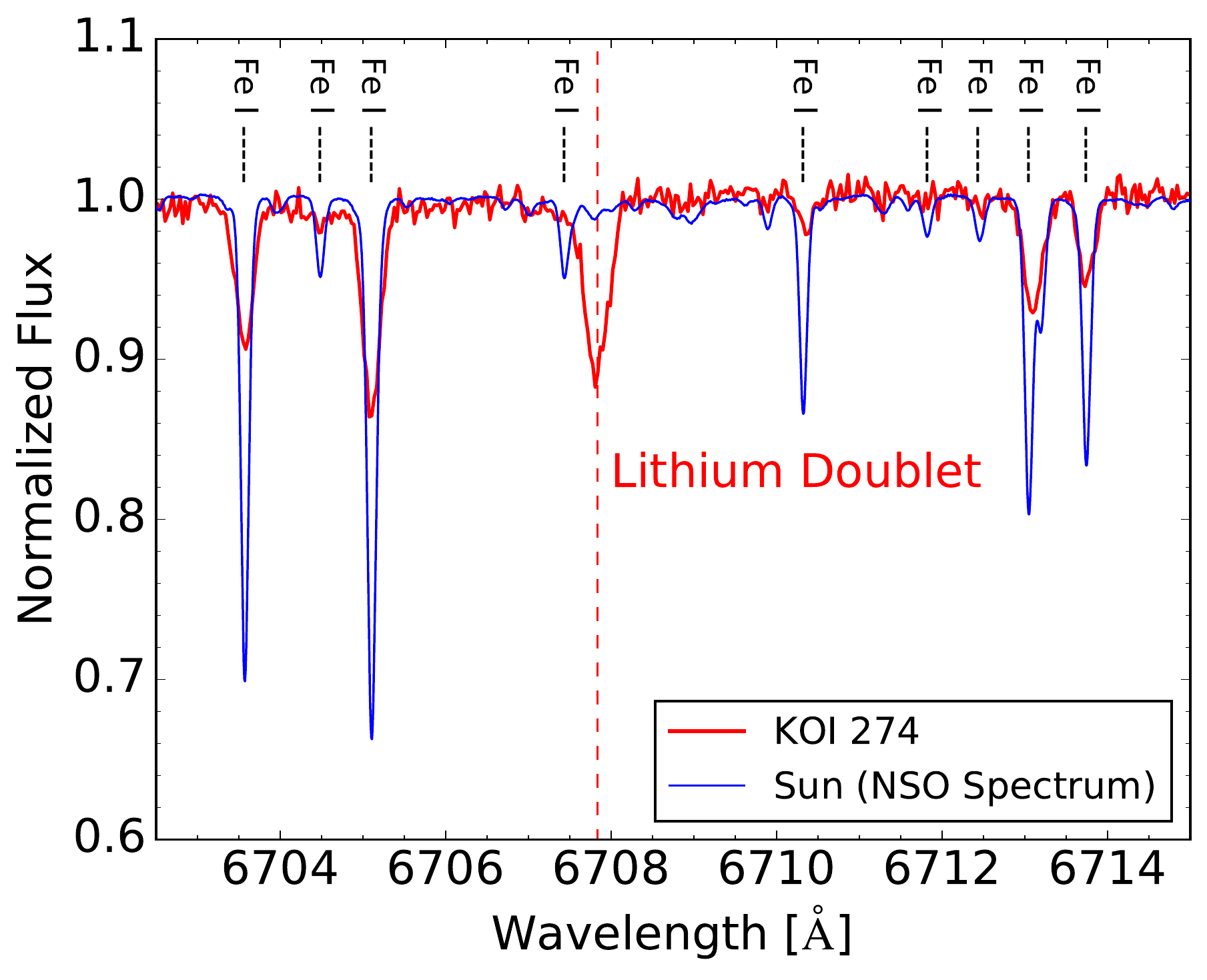}
\caption{Post-continuum normalization and wavlength calibration spectrum. In red is a HIRES spectrum of KOI 274 (\teff\ = 6081 K, \logg\ = 4.09, and [Fe/H] = --0.03), which has been continuum-normalized and wavelength-calibrated. The blue spectrum is the rest-wavelength solar spectrum from the National Solar Observatory's Solar Flux Atlas. Significant solar lines are labeled accordingly, including the Li doublet feature indicated by the red dashed vertical line.} \label{fig:compspec}
\end{figure}

Next, we applied a Doppler correction to shift the spectrum into its rest frame. We determined the Doppler correction velocity by cross-correlating the rest-wavelength, National Solar Observatory (NSO) solar spectrum with the object spectrum. The NSO spectrum is an extremely high resolution and high $S/N$ solar spectrum collected with the Brault National Solar Observatory Fourier Transform Spectrometer \citep{Hinkle2011}. Many of the same absorption lines appear in the solar and HIRES spectra due to the similar \teff\ of the Sun and our sample's stars. Therefore, we employed cross-correlation through PyRAF's $xcsao$ routine. This routine succeeded for all stars in our sample. Figure \ref{fig:compspec} displays the final product of these two routines. The example spectrum exhibits a strong Li feature, unlike the solar spectrum, but both include significant Fe lines. The difference in depth of the Fe lines results from a combination of temperature, metallicity, rotational velocity, and spectral resolution effects. Cooler, higher metallicity stars such as the Sun display stronger Fe I lines when compared to hotter, lower metallicity stars such as KOI 274, even at similar \vsini\ and spectral resolution.

\begin{figure}[htbp]
\includegraphics[width=3.35in]{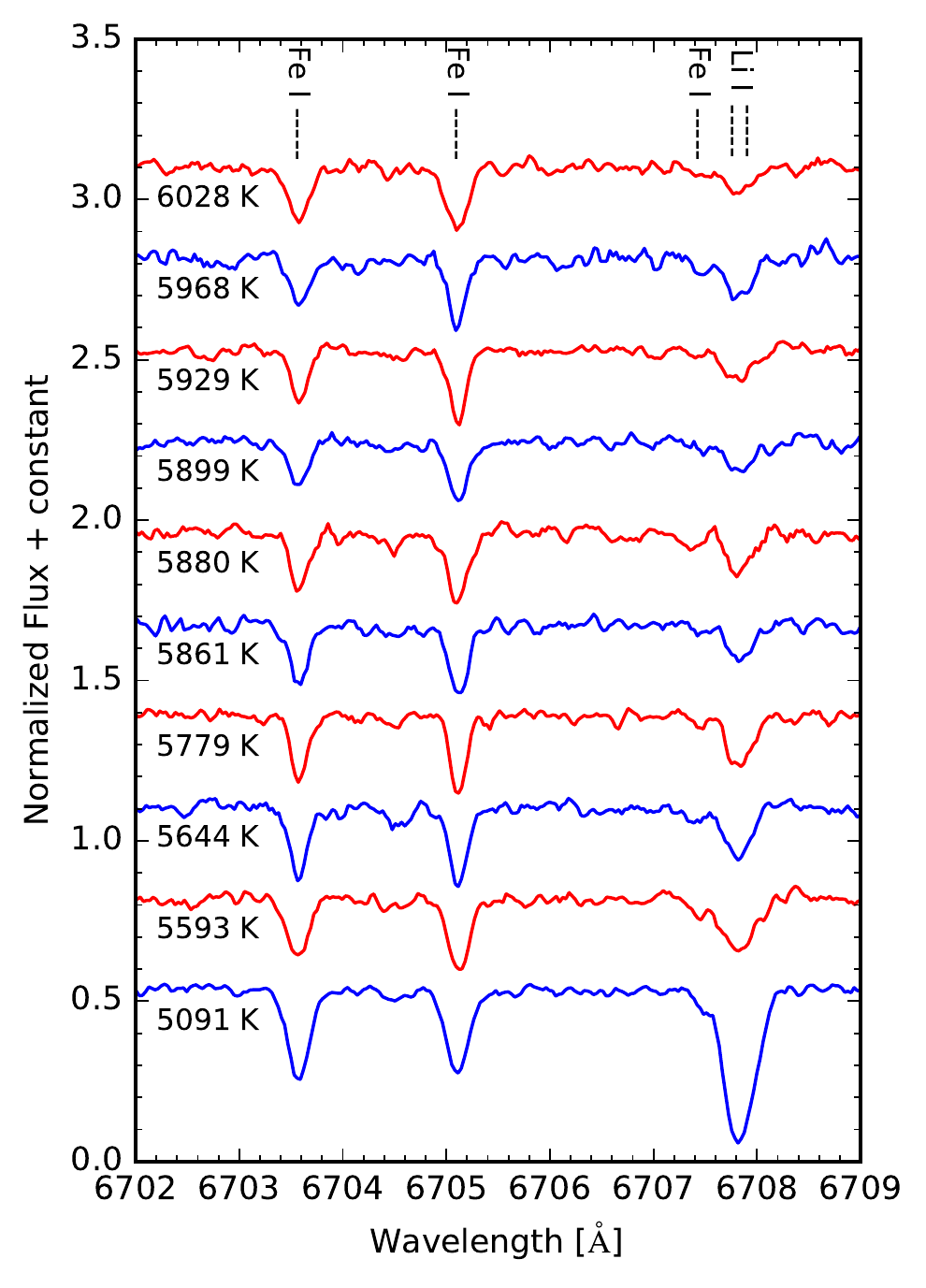}
\caption{\teff\ dependence of the Li doublet. Representative spectra at different \teff\ values are plotted here alternating between red and blue lines, all of which have been continuum normalized, wavelength calibrated, and smoothed using a three-point boxcar. These particular spectra were chosen because they have the following ranges in parameter space:  2.12 $\leq$ A(Li) $\leq$ 2.28, -0.04 $\leq$ [Fe/H] $\leq$ 0.16, and \vsini\ $<$ 6 km/s. We chose these ranges to illustrate how the Fe I lines and Li doublet change with temperature for stars of similar $S/N$, [Fe/H], A(Li), and small \vsini.} \label{fig:lidoublet}
\end{figure}

In Figure \ref{fig:lidoublet}, we illustrate the structure of spectra of multiple stars around the 6708 \AA\ Li feature for stars with a range of \teff. These particular stars were chosen because of their similar A(Li), [Fe/H], and small \vsini. As a function of \teff, stellar lithium features vary significantly in strength. Note how the Fe lines become slightly stronger as \teff\ decreases from top to bottom, while the Li feature becomes much stronger as \teff\ decreases. This illustrates the strong relationship between the Li EW and \teff. In the hotter stars, more Li is ionized, so the Li I feature weakens.

\subsection{Determining the Li EW}
Next, we measured the Li EW in the normalized and shifted spectra. The National Institute for Standards and Technology (NIST) has the Li I resonance doublet listed with one transition at 6707.76 and the other at 6707.91 \AA. We also had to account for the Fe I line that occurs at 6707.44 \AA. Because of the wide variety of spectra at different \teff, [Fe/H], and \vsini, it was difficult to find an automated EW measurement routine that was effective for all spectra in our sample.

After testing multiple automated fitting routines/packages, we concluded that Levenberg-Marquardt FIT (LMFIT) \citep{Newville2014} works well for our purposes; namely, LMFIT can simultaneously fit the Fe line and both Li lines while also providing bounds on each of the fit parameters, unlike other oversimplified methods such as numerical integration or singular Gaussian fits. LMFIT is a non-linear least-square minimization and curve fitting package for Python, which allows users to specify their own composite functions, bounds on parameters, and more. We used a four component composite model. This four component model consisted of a constant = 1 continuum level, one Gaussian for the Fe I line at 6707.44 \AA, one Gaussian for the Li I line at 6707.76 \AA, and one Gaussian for the Li I line at 6707.91 \AA.

In our model, we did not allow the continuum level to vary; we operated under the assumption our continuum normalization requires no adjustment near the Li feature. This assumption is sufficient because the vast majority of CKS stars have \vsini\ $<$ 15 km/s, in addition to all having \teff\ $>$ 4500 K. Therefore, we do not expect significant blending of lines due to the stars' small \vsini, nor significant spectral veiling from the molecular/metal absorption lines of M-dwarfs near the Li doublet. For each of the Gaussians, we implemented similar bounds on the three fitting parameters. We limited their amplitudes to [--1.0, 0.0] to prevent any positive noise fits. We limited the Gaussian widths ($\sigma$) to [0.05, 0.10] to prevent any unphysical, noise-dominated fits. Also, we bounded the set of Gaussian centers to the Fe I line center at 6707.44 \AA\ with physically required separations of 0.32 and 0.47 \AA\ for the Li I 6707.76 and 6707.91 \AA\ lines, respectively, while allowing the group as a whole to shift $\pm$ 0.06 \AA. This gives LMFIT the flexibility to shift to fit noisy line profiles but not by more than a resolution element ($\sim$0.12 \AA). Unlike other routines, LMFIT sufficiently fits Li absorption features with varying peaks and widths due to the wide range of stellar properties (\teff, \logg, \vsini, etc.) within the 1305 spectra.

\begin{figure}[t]
\includegraphics[width=3.35in]{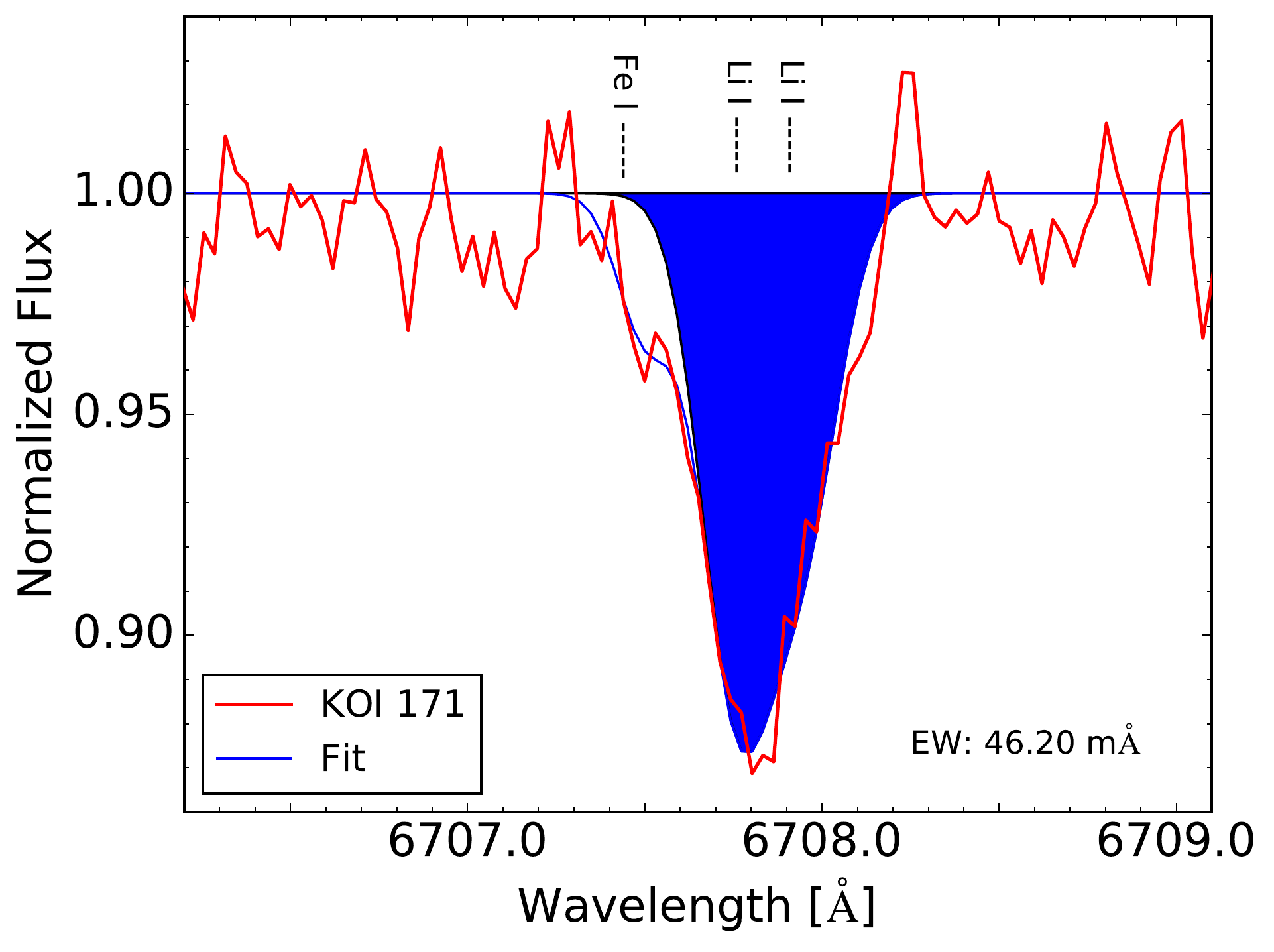}
\caption{Equivalent width determination for KOI 171 with $S/N = 42$. The red curve represents the HIRES spectrum which has been continuum normalized, Doppler-shifted, and smoothed by a three-point boxcar. The blue curve is the best fit from LMFIT's least-squares minimization process with our composite model. The blue-filled area denotes the integrated EW of the Li I doublet; the calculated EW is shown. Additionally, we indicate the locations of the Li and Fe lines.} \label{fig:ewfit}
\end{figure}

Next, we computed the Li EW. See Figure \ref{fig:ewfit} for an illustration of this method. The trough of the Fe line is not centered with respect to the Fe label. This is because the fit was improved by shifting slightly to the right. We emphasize that the calculated EWs do not include contributions from the Fe I line, as illustrated by the blue-filled area in Figure \ref{fig:ewfit}. In weak to moderate Li features like those in Figure \ref{fig:ewfit}, the feature has a slight asymmetry, caused by a difference in intensity between the smaller wavelength (greater intensity) and larger wavelength (lesser intensity) lines. This can be seen easily with very high resolution and sufficiently high $S/N$ spectra as discussed in \cite{Reddy2002}. In our fit, a slight asymmetry is present in the skewed Gaussian from the sum of the blended Li lines.

We defined the uncertainty in our EW measurement, $\sigma_{\mathrm{EW}}$, as the quadratic sum of the $S/N$-per-pixel-dependent Equation (7) ($\sigma_{\mathrm{UL}}$) in \cite{Cayrel1988}, and the average of the difference of measured EWs when modifying the continuum level $\pm$ $\frac{1}{S/N_{\mathrm{res}}}$ where $S/N_{\mathrm{res}}$ is the signal-to-noise per resolution element \citep{Lis2015}. Due to the limitations of our abundance-determination software, we report and flag our EW measurements according to the following criteria (all reported uncertainties are \sigmaew):  if the measured EW $>$ $\sigma_{\mathrm{UL}} + \sigmaeweq$, we flag the point as a Li detection and report the measured EW; if $\sigma_{\mathrm{UL}}$ $<$  EW $<$ $\sigma_{\mathrm{UL}} + \sigmaeweq$, we report the measured EW but flag the point as an upper limit; if the measured EW $<$ $\sigma_{\mathrm{UL}}$, we report EW = $\sigma_{\mathrm{UL}}$ and flag the point as an upper limit. See Table \ref{tab:full} for the entire sample's reported EWs.

\begin{figure}[t]
\includegraphics[width=3.35in]{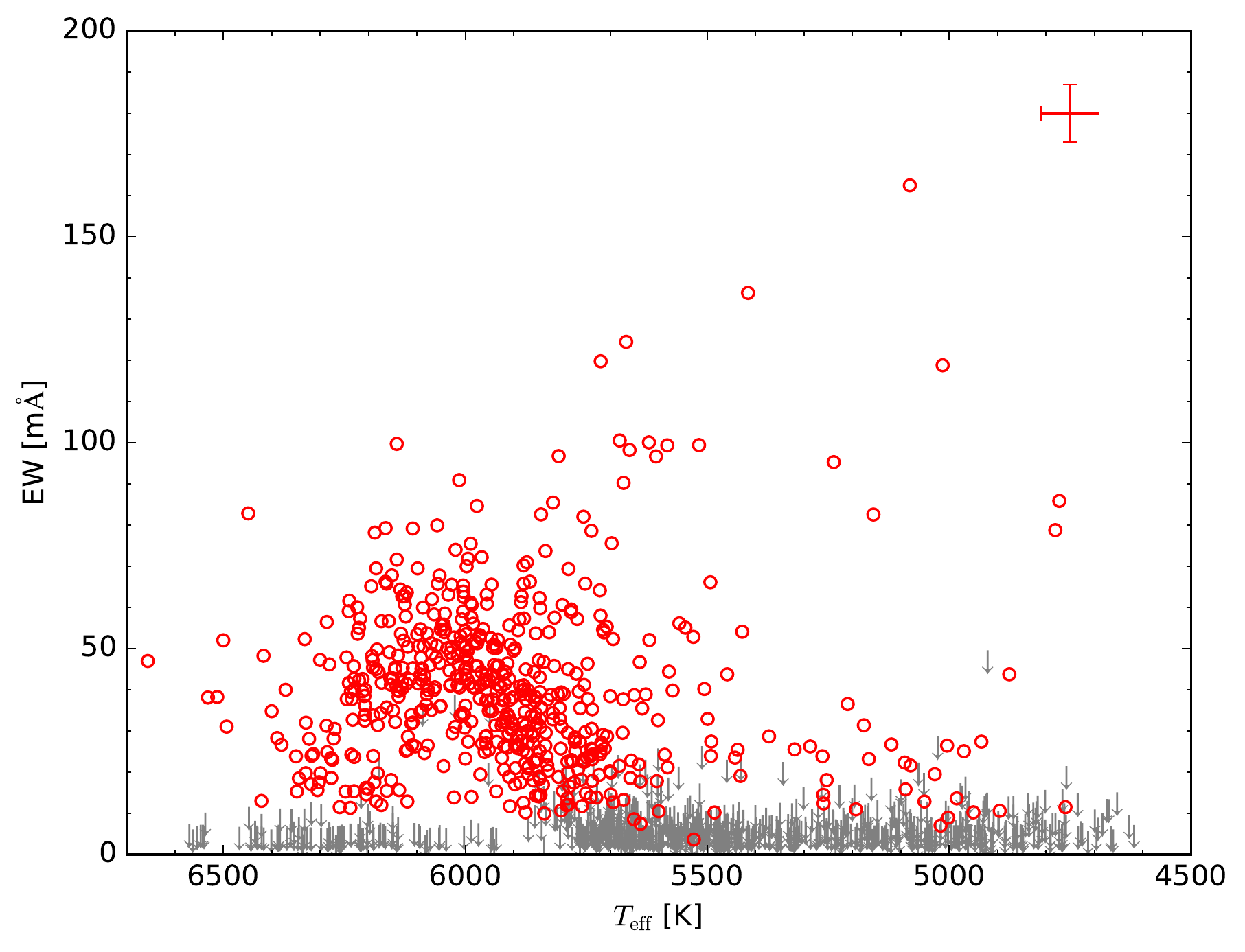}
\caption{Li EW as a function of effective temperature for all CKS stars. The red points represent stars with Li detections, while the grey downward arrows are Li EW upper limits (EW $<$ $\sigma_{\mathrm{UL}} + \sigmaeweq$). Typical error bars are supplied in the upper right corner of the plot for reference.} \label{fig:ewteff}
\end{figure}

In Figure \ref{fig:ewteff}, we plot the measured EW of all CKS stars as a function of \teff. The upper limits (grey downward arrows) are stars with measured Li EW $<$ $\sigma_{\mathrm{UL}} + \sigmaeweq$. From this plot, we can identify young stars:  those with large EWs at low \teff. Any $Kepler$ planet host stars with EWs located far above the ``slipper'' are particularly young. At higher \teff, the slipper is less well-defined largely because we do not have as many of these larger stars, and those that we do have are close to the Li ``dip'' observed in the Hyades as discussed in \cite{Boesgaard2016}.

\subsection{Model Atmosphere Interpolation}
We utilized Model Atmosphere in Radiative and Convective Scheme (MARCS) \citep{Gustafsson2008} model atmospheres to convert EW to A(Li). Unlike Kurucz model grids, MARCS grids include the microturbulence parameter ($\xi$) in addition to \teff, \logg, and [Fe/H]. In particular, we chose MARCS plane-parallel grids because our sample is primarily composed of main sequence dwarfs. We adopted the microturbulent description of Equations (1) and (2) in \cite{Takeda2013}. We then interpolated from the discrete MARCS grids to the model atmospheres representing the adopted CKS stellar parameters. 

\begin{figure}[t]
\includegraphics[width=3.35in]{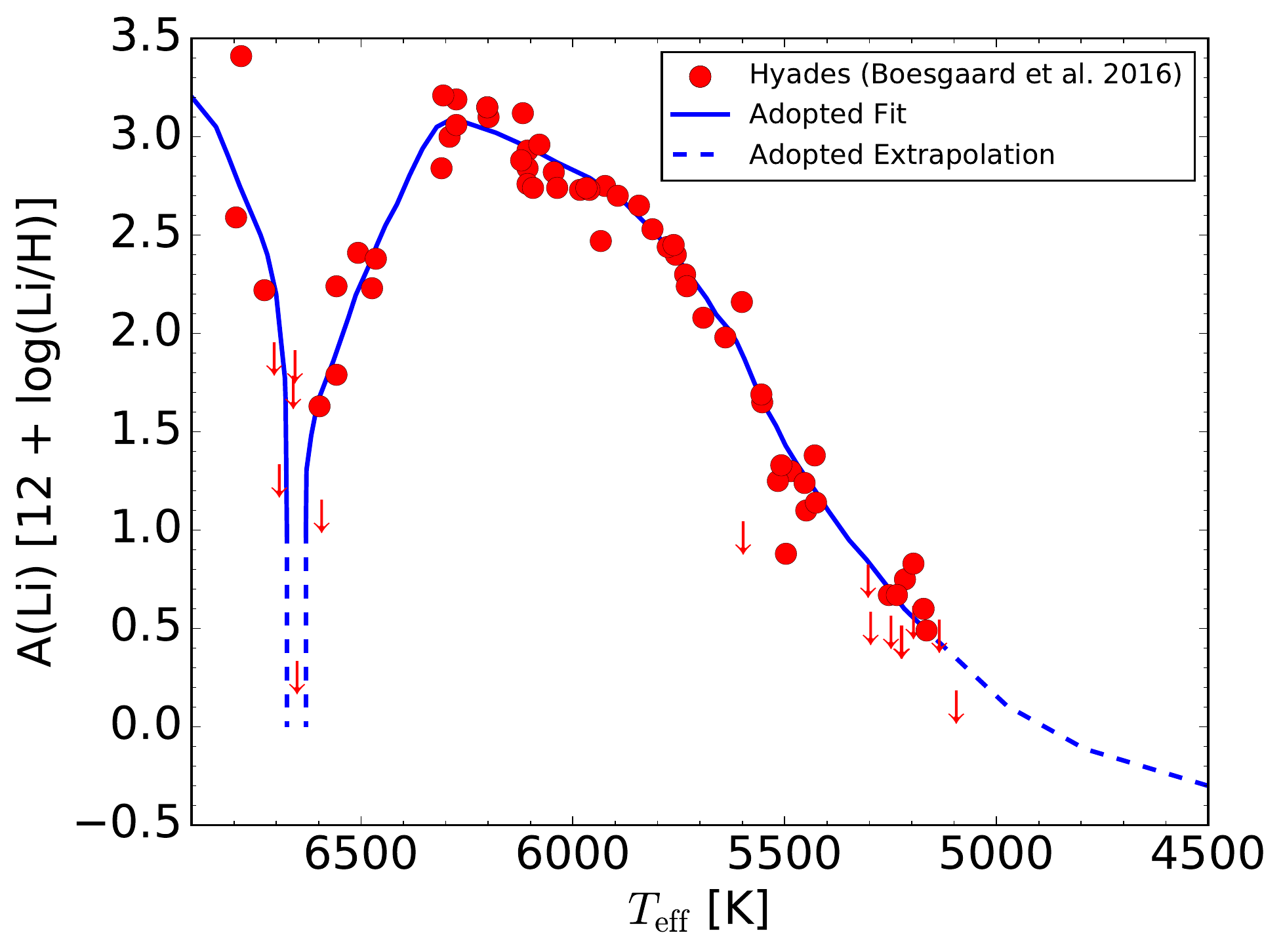}
\caption{Empirical A(Li) versus \teff\ curve for the Hyades. The red points are Hyades data from \cite{Boesgaard2016}, where the downward arrows signify upper limits. The blue curve is the approximate fit to this data. The dashed portions of the curve represent regions either where we extrapolated (\teff\ $<$ 5100 K) or where we have only upper limits (\teff\ $\sim$6650 K).}\label{fig:hyades}
\end{figure}

\begin{figure*}[htbp]
\includegraphics[width=7in]{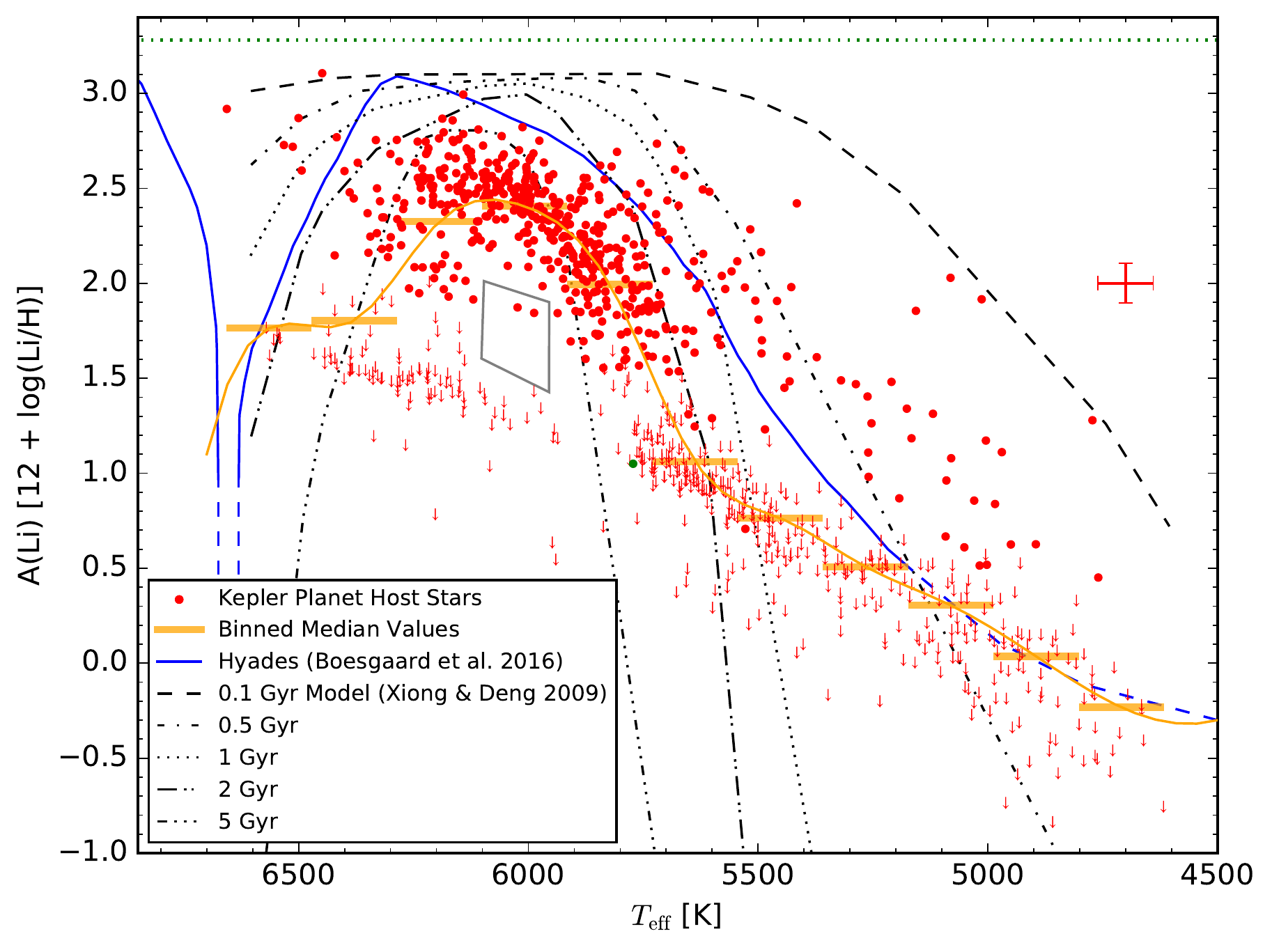}
\caption{A(Li) as a function of \teff\ for 1025 high $S/N$ ($>$ 30) $Kepler$ planet host stars (in red). Downward arrows represent upper limits, while circles are spectra with $EW_{\mathrm{Li}}$ $>$ $\sigma_{\mathrm{UL}} + \sigmaeweq$. The orange horizontal bars show the binned median abundances for each of the temperature bins that are 184 K wide and include the upper limits. The orange curve is a cubic spline interpolation between the binned median abundances. The blue curve represents an approximate fit of the Hyades from \cite{Boesgaard2016}; the dashed blue line at \teff\ $\approx$ 6600 K illustrates the Li ``dip'' where only upper limits have been measured, while the dashed blue line at \teff\ $<$ 5100 K is our adopted extrapolation. The dashed/dotted black lines are from \cite{Xiong2009} and represent theoretical model isochrones for Li depletion in MS stars. The solid grey polygon at \teff\ $\approx$ 6000 K and A(Li) $\approx$ 1.8 is the Li desert illustrated in \cite{Ramirez2012}. We plot the meteoric A(Li) = 3.28 $\pm$ 0.05 \citep{Lodders2009} and photospheric A(Li) = 1.05 $\pm$ 0.10 \citep{Asplund2009} as the green dotted line and circle, respectively. The red error bars show the sample's median errors in A(Li) and \teff. Because A(Li) depends sensitively on \teff, we stress that the errors are correlated.} \label{fig:aliteff}
\end{figure*}

\begin{figure*}[htbp]
\includegraphics[width=7in]{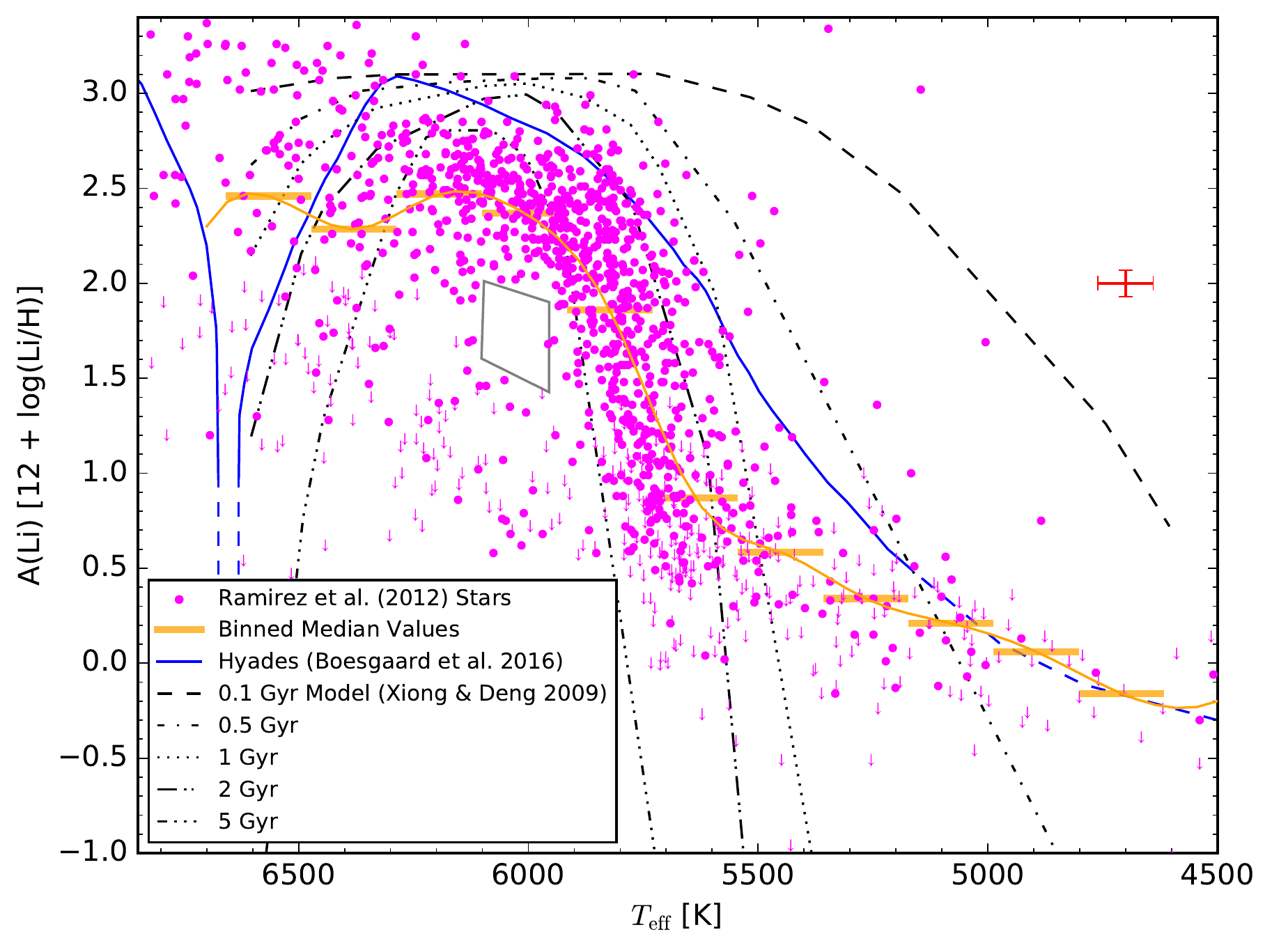}
\caption{A(Li) as a function of \teff\ for 1381 main sequence dwarfs and subgiant stars from \cite{Ramirez2012}. We provide this plot as a comparison to the CKS sample. Additionally, we have condensed the natural axes in both \teff\ and A(Li) to match Figure \ref{fig:aliteff}. As a result, there are $\sim$20 data points outside the range of the chosen axes. Downward arrows represent upper limits, while circles are spectra with measured Li EWs. The orange horizontal bars show the binned median abundances for each of the temperature bins that are 184 K wide as for our sample in Figure \ref{fig:aliteff}. The orange fit is a cubic spline interpolation between the binned median abundances. The blue curve represents an approximate fit of the Hyades from \cite{Boesgaard2016}; the dashed blue line illustrates the Li ``dip'' at \teff\ = 6600 K where only upper limits have been measured, while the dashed blue line at \teff\ $<$ 5100 K is our adopted extrapolation. The dashed/dotted black lines are from \cite{Xiong2009} and represent theoretical model isochrones for Li depletion in MS stars. The solid grey polygon at \teff\ $\approx$ 6000 K and A(Li) $\approx$ 1.8 is the Li desert illustrated in \cite{Ramirez2012}. The red error bars show the sample's median errors in A(Li) and \teff.} \label{fig:ramirezaliteff}
\end{figure*}

\subsection{Computing the Lithium Abundance}
We determined A(Li) using MOOG \citep{Sneden2012}. This code performs a variety of LTE analysis and spectrum synthesis tasks. We used its $blends$ routine, which computes A(Li). $Blends$ fits abundances of species by using a given model atmosphere to match blended-line EWs. We utilized $^7\mathrm{Li}$ hyperfine splitting transition wavelengths \citep{Sansonetti1995} and $gf$ values \citep{Yan1995} from Table (3), adapted from \cite{Andersen1984}, in \cite{Smith1998} as our line list. We did not include the nearby Fe I line within our line list because we only computed the Li feature's EW using LMFIT. We then applied $blends$ to determine A(Li) using this line list, in addition to the Li EW and interpolated model atmosphere from \S3.2 and \S3.3, respectively.

We calculated the uncertainty, $\sigma_{\mathrm{A(Li)}}$, using a similar method to \cite{Ramirez2012}. First, we varied each of the MOOG input parameters individually (\teff, \logg, [Fe/H], and $EW_{\mathrm{Li}}$) according to their internal CKS 1$\sigma$ errors and then recalculated A(Li). This resulted in two Li abundances, one corresponding to the stellar model after a 1$\sigma$ increase in the varied stellar parameter, A(Li)$_+$, and the other corresponding to the stellar model after a 1$\sigma$ decrease in the varied stellar parameter, A(Li)$_-$. Next, we calculated the largest deviation of the upper and lower bound A(Li) values from the A(Li) corresponding to the adopted parameter. We repeated this process for the rest of the MOOG input parameters, and then added the largest deviations of each input parameter in quadrature to determine $\sigma_{\mathrm{A(Li)}}$.

As \cite{Lis2015} discussed in their Appendix, adding separate errors in quadrature is insufficient because of the nonlinear transformation between stellar parameters/equivalent widths and abundances. Therefore, our reported A(Li) errors are quantitatively incorrect. However, because we use the largest deviations from A(Li) as our adopted individual uncertainties and then add them in quadrature, we posit that we overestimate the true abundance errors on one side due to the asymmetric distribution of abundances. To appropriately determine each $\sigma_{\mathrm{A(Li)}}$, we would need to perform a Markov Chain Monte Carlo (MCMC) error analysis, which would require an extreme amount of computing time (MOOG would need to be run $>$ 1000 times per star) for the 1305 stars in our sample. In reality, the errors in the individual A(Li) are not particularly important for the results of this paper. Consequently, we conservatively estimate $\sigma_{\mathrm{A(Li)}}$ in the symmetric manner described above.

To ensure the accuracy and precision of MOOG's $blends$ routine, we compared our EW-A(Li) results to spectral synthesis A(Li) using the same stellar parameters and the Li-blend EW for a subset of 18 stars. We found the measurements to be consistent within 4$\%$.

\section{Lithium Abundances}
In Figure \ref{fig:hyades}, we plot the empirical A(Li) versus \teff\ curve for the Hyades based on data from \cite{Boesgaard2016}. We ignored all upper limits while constructing this curve. To construct the Hyades curve, we performed manual linear interpolation of A(Li) as a function of \teff, dictated by the location of individual Hyades stars in this plot. The dashed portions of the curve indicate two separate extrapolations:  a) at \teff\ $<$ 5100 K, following the smooth curve of the interpolation at higher \teff\ and flattening to the upper limits of the CKS sample towards \teff\ = 4500 K, and b) at \teff\ $\approx$ 6650 K, where only upper limits exist, hence the vertical dropoff. At \teff\ $>$ 6800 K, the curve is influenced heavily by a few data points outside the range of this plot.

The blue Hyades curve is important when viewing Figure \ref{fig:aliteff}, as it provides an empirical relationship between A(Li), \teff, and age, much like the theoretical isochrones (black, dashed and dotted curves) from \cite{Xiong2009} do. In Figure \ref{fig:aliteff}, the red circles and downward arrows represent detected and upper limit A(Li) values, respectively, for all 1025 CKS $Kepler$ planet host stars that have $S/N$ $>$ 30 spectra. We emphasize that the \teff\ and A(Li) errors are correlated, so when we change \teff, A(Li) will be affected as well.

Table \ref{tab:full} contains our entire catalog of A(Li) measurements, including each observation code, KOI number, $S/N$, adopted \teff, adopted \logg, adopted [Fe/H], the calculated $\xi$, the measured EW and its uncertainty, and the computed $\sigma_{\mathrm{A(Li)}}$. The entire table, in machine-readable format, can be found in the online version of this journal.

\subsection{Identification of Young Stars}
Unfortunately, deriving precise ages from measurements of $EW_{\mathrm{Li}}$ and subsequent computation of A(Li) is quite difficult due to the inability of current models \citep{Xiong2009} to fit observed abundances. In Figure \ref{fig:aliteff}, the distribution of $Kepler$ planet host stars resembles that of the Hyades much better than these Li depletion model isochrones. The orange bars represent the binned median A(Li) for stars with \teff\ spanning the width of the bar. The orange curve is a cubic spline interpolation of the orange median A(Li) bars and serves as a statistical representation of the median A(Li) stars at each effective temperature. These stars may represent an empirical isochrone, much like the Hyades do. This curve appears similar in shape to the Hyades curve, while it intersects multiple theoretical model isochrones from \cite{Xiong2009}. Because the theoretical models are unable to match both the Hyades and the distribution of $Kepler$ planet host stars, the ages indicated by each of the black curves prove unreliable. Although numerical ages are difficult to determine, we can use A(Li) versus \teff\ plots to distinguish between young (i.e. $<$ 650 Myr) and old systems.

\begin{figure*}[htbp]
\includegraphics[width=7in]{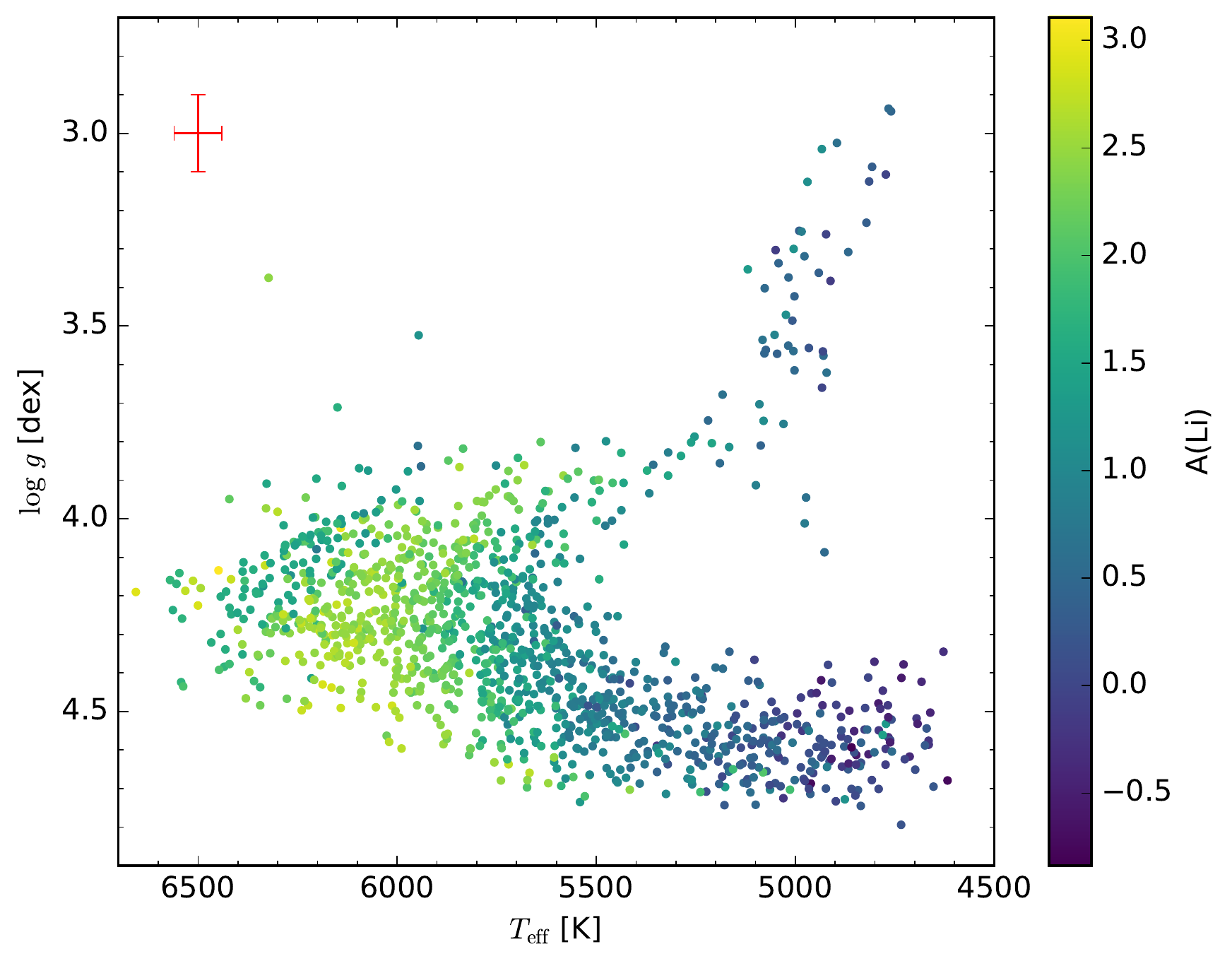}
\caption{Hertzprung-Russell diagram of \logg\ as a function of \teff\ for all 1305 $Kepler$ planet host stars. The red error bars show the sample's median errors in \logg\ and \teff. The color of the points represents A(Li) on a linear scale as illustrated by the color bar on the right.} \label{fig:loggteffali}
\end{figure*}

Stars deplete their surface Li over time. However, the rate of depletion varies with \teff. Cooler stars (K type and later) deplete their Li faster because their convective zone depths are larger than hotter (G-type and earlier) stars. Therefore, we expect to see more stars with high A(Li) at higher \teff, as demonstrated in Figure \ref{fig:aliteff}. Naturally, we see a higher proportion of upper limits at cool temperatures compared to hot temperatures. We expect younger stars to have higher A(Li) than other stars at their \teff. With this in mind, we can pick out the youngest stars as those most significantly above the blue Hyades curve and our orange empirical median curve.

Additionally, we can define a sample of stars that are younger than the Hyades by computing:
\begin{equation}
\Delta_{\mathrm{A(Li),Hyades}} = \mathrm{A(Li)} - \mathrm{A(Li)_{Hyades}}
\end{equation}
for each star in the sample, where A(Li) is the computed value from the pipeline and A(Li)$_{\mathrm{Hyades}}$ is the interpolated value of the Hyades at the star's \teff. Stars that are younger than the Hyades will have $\Delta_{\mathrm{A(Li),Hyades}}$ $>$ 0. Table \ref{tab:youngstars} includes all young CKS stars with detected Li (no upper limits are included in the table). The subtraction of the empirical Hyades curve is useful because it removes the offset caused by the \teff\ dependence of A(Li).

Figure \ref{fig:aliteff} indicates that stars with A(Li) $>$ 1.0 and \teff\ $<$ 5300 K are the youngest systems due to their unusually high A(Li). For comparison, the Hyades is 650 Myr according to estimates from \cite{Perryman1998} and \cite{Brandt2015}. Therefore, stars above the blue Hyades curve should be younger than 650 Myr. In essence, we can determine the relative ages of systems in Figure \ref{fig:aliteff} by subtracting the blue Hyades curve from the $Kepler$ planet host star data points. These A(Li)-\teff\ plots also allow the qualitative comparison of average system properties above and below the Hyades.

For a comparison to our CKS sample, we include Figure \ref{fig:ramirezaliteff}, which contains data from \cite{Ramirez2012}. We plot the same structures as those in Figure \ref{fig:aliteff}. In \cite{Ramirez2012}, the Li desert (outlined as a grey trapezoid in Figures \ref{fig:aliteff} and \ref{fig:ramirezaliteff}) was emphasized as an area without stars. The authors argued that the Li desert is a physical phenomenon caused by short-lived processes on the stellar surface that deplete Li for 1.1--1.3 $M_{\odot}$ stars. These processes are not well understood, but the observational evidence is hard to ignore. However, our sample produces two stars within this desert. If the Li desert is indeed a physical gap, we conclude that errors in both A(Li) and \teff\ can account for this discrepancy. Moreover, Figure \ref{fig:aliteff} (our sample) has a large number of upper limits when compared to Figure \ref{fig:ramirezaliteff} \citep{Ramirez2012}. This is due to our low median $S/N \approx 45$ compared to the median $S/N \gtrapprox 100$ from \cite{Ramirez2012}. Figure \ref{fig:aliteff} has more cool stars than Figure \ref{fig:ramirezaliteff}, while Figure \ref{fig:ramirezaliteff} has more hot stars. This is an important distinction between the two samples. Faint stars in the CKS sample were chosen because of higher planet multiplicity and/or the presence of interesting planets. Because of this, we have a larger fraction of low \teff\ stars compared to \cite{Ramirez2012}. Curiously, only one star appears above the Hyades at \teff\ $\approx$ 5900--6300 K in Figure \ref{fig:aliteff}, while Figure \ref{fig:ramirezaliteff} has $\gtrsim$ 10 stars in the same area. Based on the simple assumption of a uniform distribution of stellar ages, Figure \ref{fig:aliteff} should have a few stars above the Hyades at these \teff. This observation puzzles us.

We are also puzzled by the large number of stars within the Li dip in Figure \ref{fig:ramirezaliteff}. According to the Li depletion mechanisms discussed in \cite{Xiong2009}, the Li dip is a result of gravitational settling of Li into progressively hotter radiative zones in the stellar interior, where it is burned in ($p$,$\alpha$) reactions. Why so many stars from \cite{Ramirez2012} reside in the dip perplexes us. We do not see the same within the CKS sample, but we do not have many stars in that range of \teff.

Table \ref{tab:young} displays the full list of the CKS planets with significant Li detections that are younger than the Hyades. However, we do not incorporate stars with upper limits because their A(Li) are unreliable, nor do we include false positive planet detections in this table.

\subsection{Stellar Properties and Age}

\begin{figure*}[htbp]
\includegraphics[scale=.74]{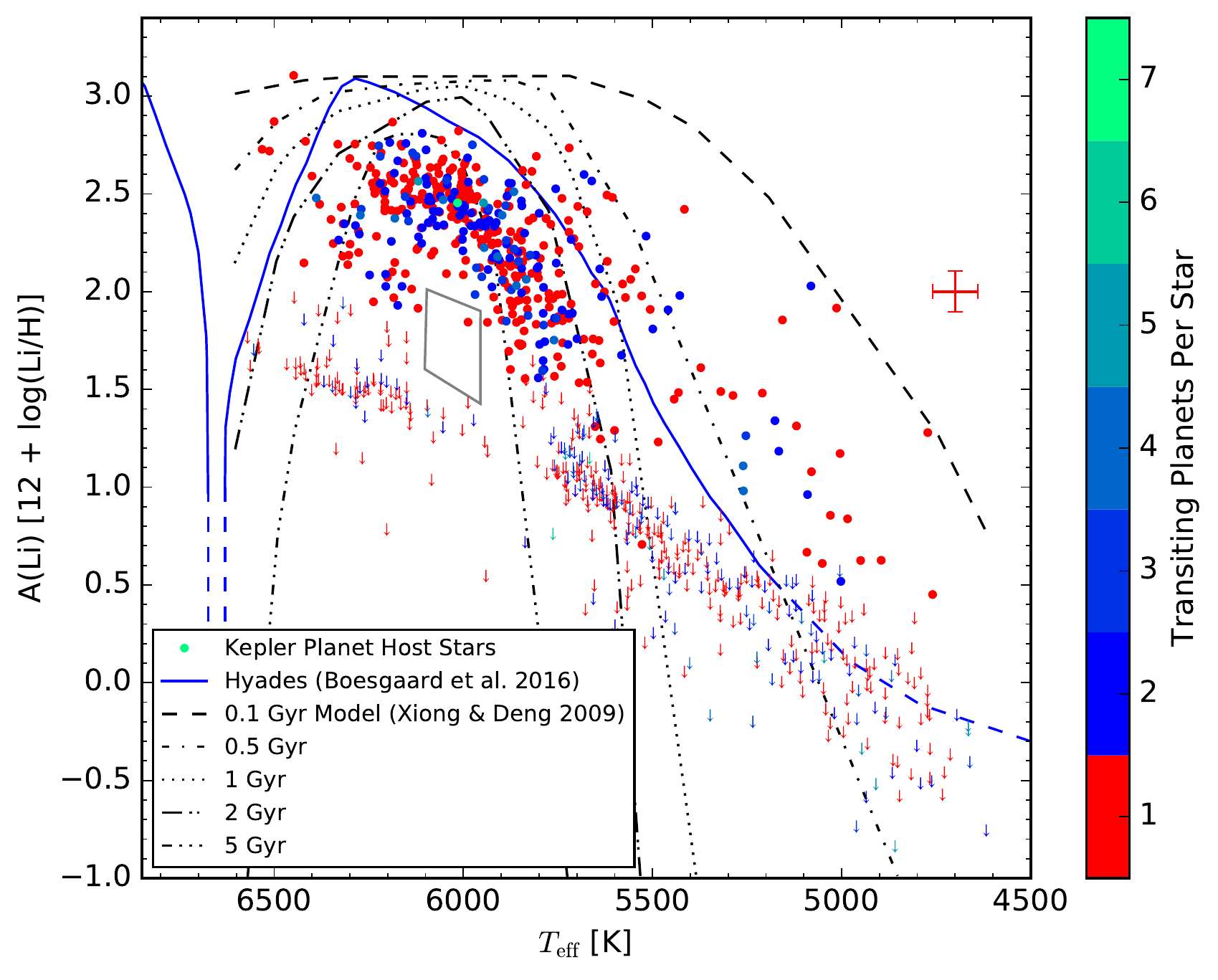}
\centering
\caption{A(Li) as a function of \teff\ for 918 $Kepler$ planet host stars. The data in this figure is similar to Figure \ref{fig:aliteff}, except all single star planet false positives have been removed. Downward arrows represent upper limits, while circles are stars with $EW_{\mathrm{Li}}$ $>$ $\sigma_{\mathrm{UL}} + \sigmaeweq$. The color of the points represents the number of discovered transiting planets in each $Kepler$ system as shown by the discrete color bar on the right.} \label{fig:aliteffmult}
\end{figure*}

\begin{figure*}[htbp]
\includegraphics[scale=.74]{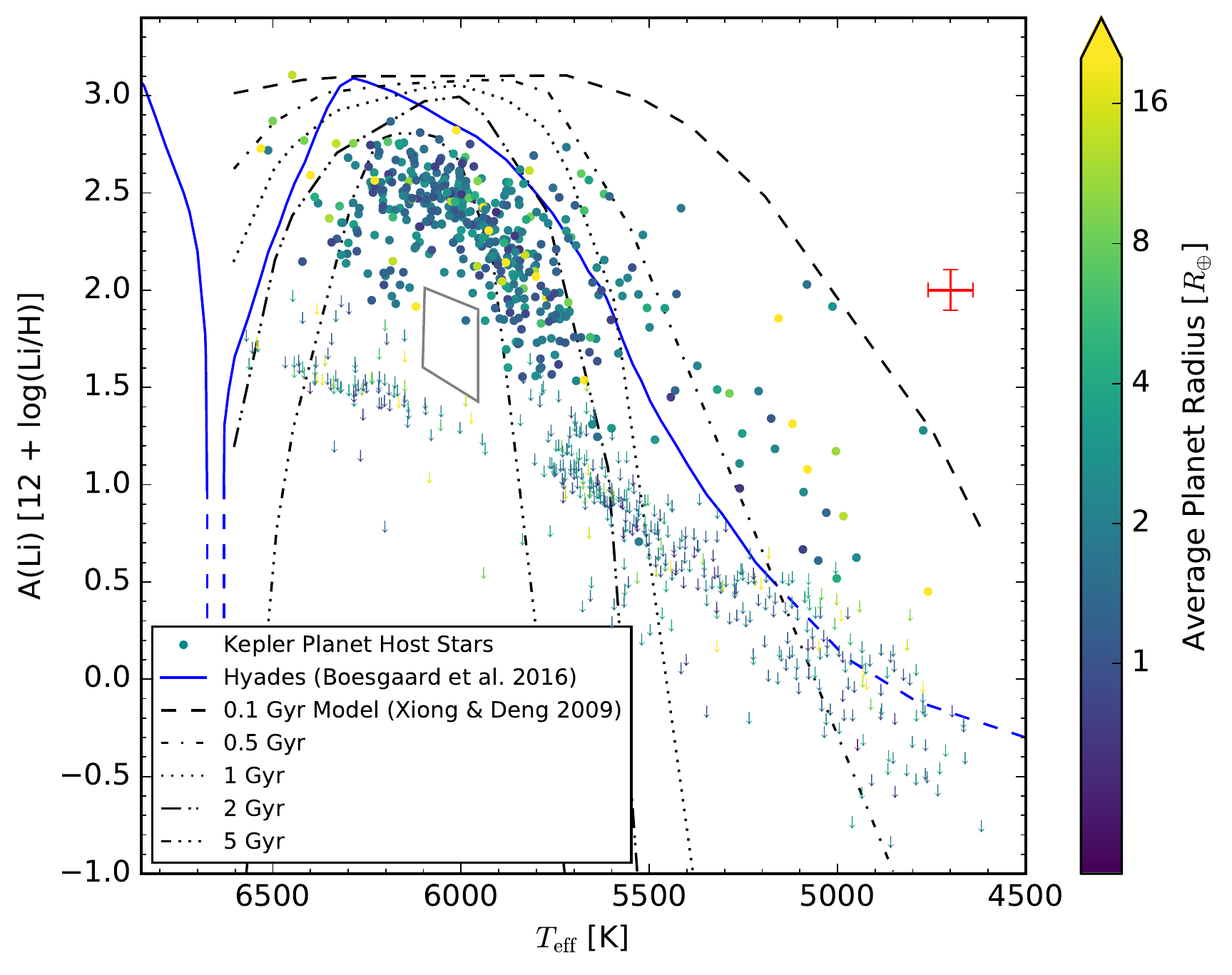}
\centering
\caption{Same as Figure \ref{fig:aliteffmult}, except colors now represent the average planet radius for each system on a continuous, logarithmic scale as shown by the color bar.} \label{fig:aliteffprad}
\end{figure*}

We begin our age investigation by analyzing trends in stellar properties with A(Li). We again utilize the A(Li) versus \teff\ plot much like Figures \ref{fig:aliteff} and \ref{fig:ramirezaliteff}, but color the points according to the stellar property of interest. First, we investigate [Fe/H]. We find no significant trends in metallicity with age after comparing stars with A(Li) above and below the Hyades, although a clump of low metallicity points occurs both above and immediately to the right of the Li desert. We arrive at similar conclusions for \logg\, although there do appear to be some ``young'' subgiants/giants around \teff\ = 5000 K and A(Li) = 1.0. This will become important for clean sample selection later. In addition, the stars with the highest A(Li) at their respective \teff\ have high \logg.

\begin{figure*}[htb]
\includegraphics[width=7.0in]{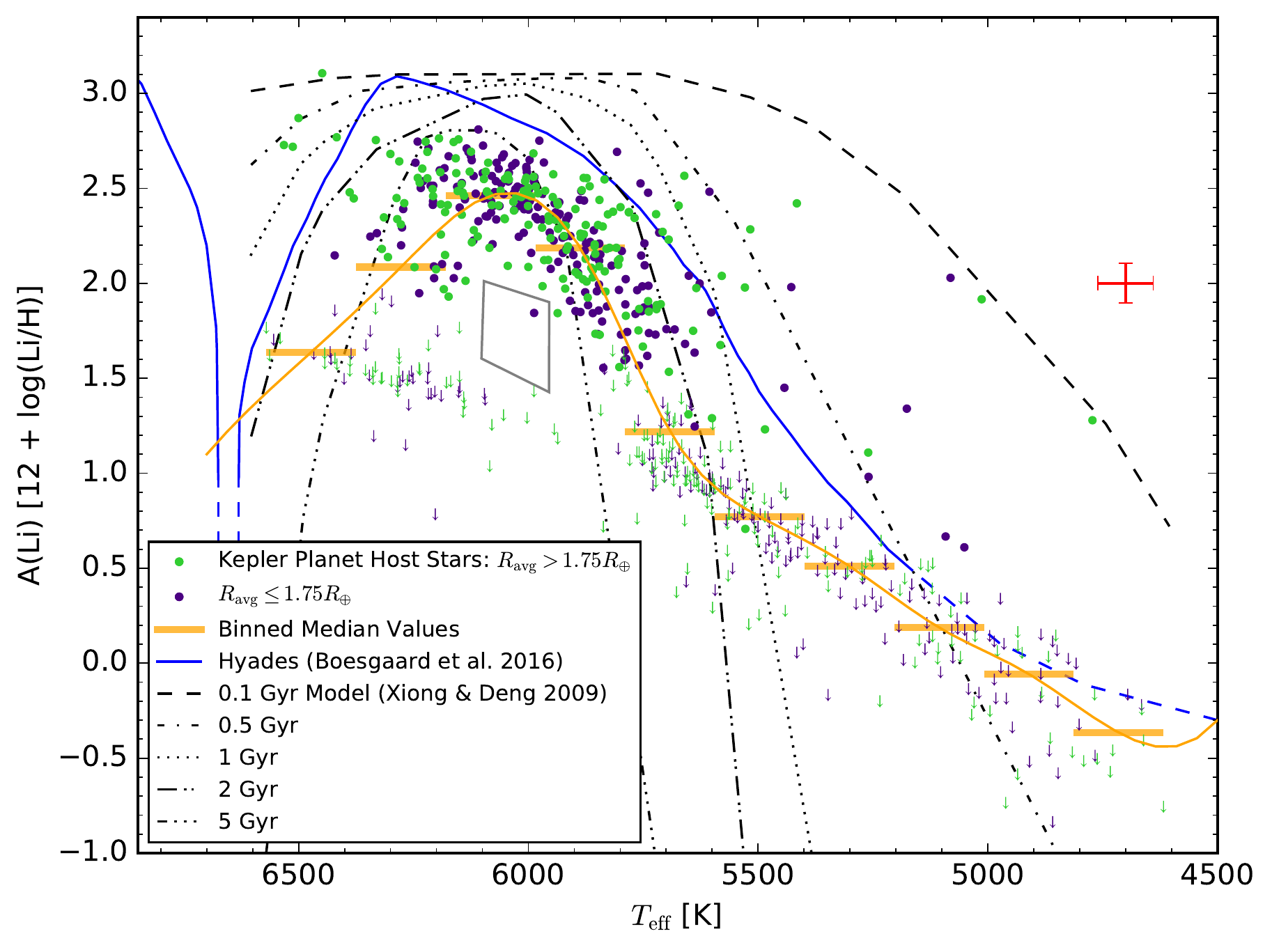}
\caption{Same axes and labels as Figure \ref{fig:aliteffprad}. From the data in Figure \ref{fig:aliteffprad}, we removed all planets with high impact parameters, those orbiting subgiant and giant stars, and those with low transit signal-to-noise ratios ($S/N$ $<$ 10). Points are colored according to whether the average radius of the planets orbiting that host star is greater than (green) or less than (purple) 1.75$R_{\oplus}$. There are 363 green points and 371 purple points.} \label{fig:aliteffpradcomp}
\end{figure*}

Figure \ref{fig:loggteffali} is a proxy for a Hertzprung-Russell (HR) Diagram with \logg\ versus \teff\ and points colored by their A(Li). The youngest systems are the brightest (green and yellow colored) points on the main sequence at their respective \teff\ and \logg. Most stars in the sample are main sequence dwarfs, although the sample also includes the horizontal branch of subgiants and then giants at the top of the ``tail'' at the lowest \teff\ and \logg. In addition, this plot reveals A(Li)'s temperature dependence, visible in the smooth transition of colors from hotter to cooler effective temperatures. We note that a few stars on the lower envelope of the main sequence in the figure (at the highest \logg) typically have larger A(Li) compared to stars at the same \teff\ and slightly lower \logg. Current stellar evolution models predict that, as stars evolve during their main sequence lifetimes, they gradually increase in luminosity and inflate in size. Therefore, the stars with the highest \logg\ values at their respective \teff\ are likely some of the youngest stars in our sample.

\subsection{Finding and Comparing Exoplanet Properties}
\footnotetext{Accessed 7/15/17}
Just as we investigated stellar parameters versus age in the previous section, we can apply the same analysis to exoplanet parameters. We use the catalog of exoplanet parameters provided by \cite{Johnson2017} and the NASA Exoplanet Archive\altaffilmark{2} to obtain exoplanet parameters for our systems. We note that many of the NASA Exoplanet Archive planet parameters derive from the detailed stellar analysis of \cite{Huber2014}. First, we investigate whether planet multiplicity (number of planets discovered per star) varies with age. In Figure \ref{fig:aliteffmult}, single planet systems are colored red, while multi-planet systems range from blue (2 planets) through bright green (7 planets). We compare points above and below the Hyades, and we find no evidence for multiplicity's dependence on age. Similarly, we investigated whether there are any trends in planet disposition, average planet period, and average planet insolation flux using more A(Li) versus \teff\ plots, but no patterns were apparent. Therefore, we conclude that these exoplanetary properties show no dependence on age.

Finally, we consider the average planet radius in each system. A quick look at Figure \ref{fig:aliteffprad} does not reveal any clear trends between planet radius and location above/below the Hyades. However, many of the identified young systems in \S4.1 (those at \teff\ $<$ 5500 K and above the Hyades curve) are green and yellow-colored. Therefore, they have large average planet radii. We find this evidence interesting, as planets are expected to deflate as they age and their star remains on the main sequence \citep{Lopez2012}. We continue with a more thorough investigation.

\subsection{Planet Radius and Age}
Before we proceed, we must consider sample biases and eliminate any systems which may introduce biases in age and planet size. Therefore, we utilize similar cuts to produce a clean sample as in \cite{Fulton2017}. First, we eliminate all false positives identified in the CKS \citep{Petigura2017} and stars with spectral $S/N$ $<$ 30. Next, we remove all planets with impact parameter, $b$, $>$ 0.7 and those orbiting subgiant and giant stars according to Equation (1) in \cite{Fulton2017}. Additionally, we remove all planets with low transit signal-to-noise ratios ($S/N$ $<$ 10). These cuts ensure that the planetary radii are reliable and that the host stars are main sequence dwarfs (where A(Li) is a reliable indicator of age). We do not remove planets with long orbital periods and large KepMags because these cuts further reduce our sample's size, while the likelihood of systematic biases in age and planetary radii of these systems is small.

To make any obscured trends more apparent in Figure \ref{fig:aliteffprad}, we split the sample into two parts:  systems with $R_{\mathrm{avg}} > 1.75R_{\oplus}$ and those with $R_{\mathrm{avg}} \leq 1.75R_{\oplus}$. We choose 1.75$R_{\oplus}$ as our separating radius because it corresponds to the trough of the empirical gap revealed in \cite{Fulton2017}. Figure \ref{fig:aliteffpradcomp} displays our ``clean'' sample in another A(Li)--\teff\ comparison plot with the points colored according to their average planet radius.

We observe from Figure \ref{fig:aliteffpradcomp} that the stars younger than the Hyades typically have planet companions on the large side of the planet radius gap discovered by \cite{Fulton2017}. The prevalence of green points above the Hyades supports the conclusion that younger planets are larger than older planets. However, because this plot assigns an average planet radius to each star, we lose information about individual planetary radii. Therefore, we must perform additional statistical analysis to test our hypothesis that, on average, larger planets orbit younger stars.

\begin{figure}[t]
\includegraphics[width=3.35in]{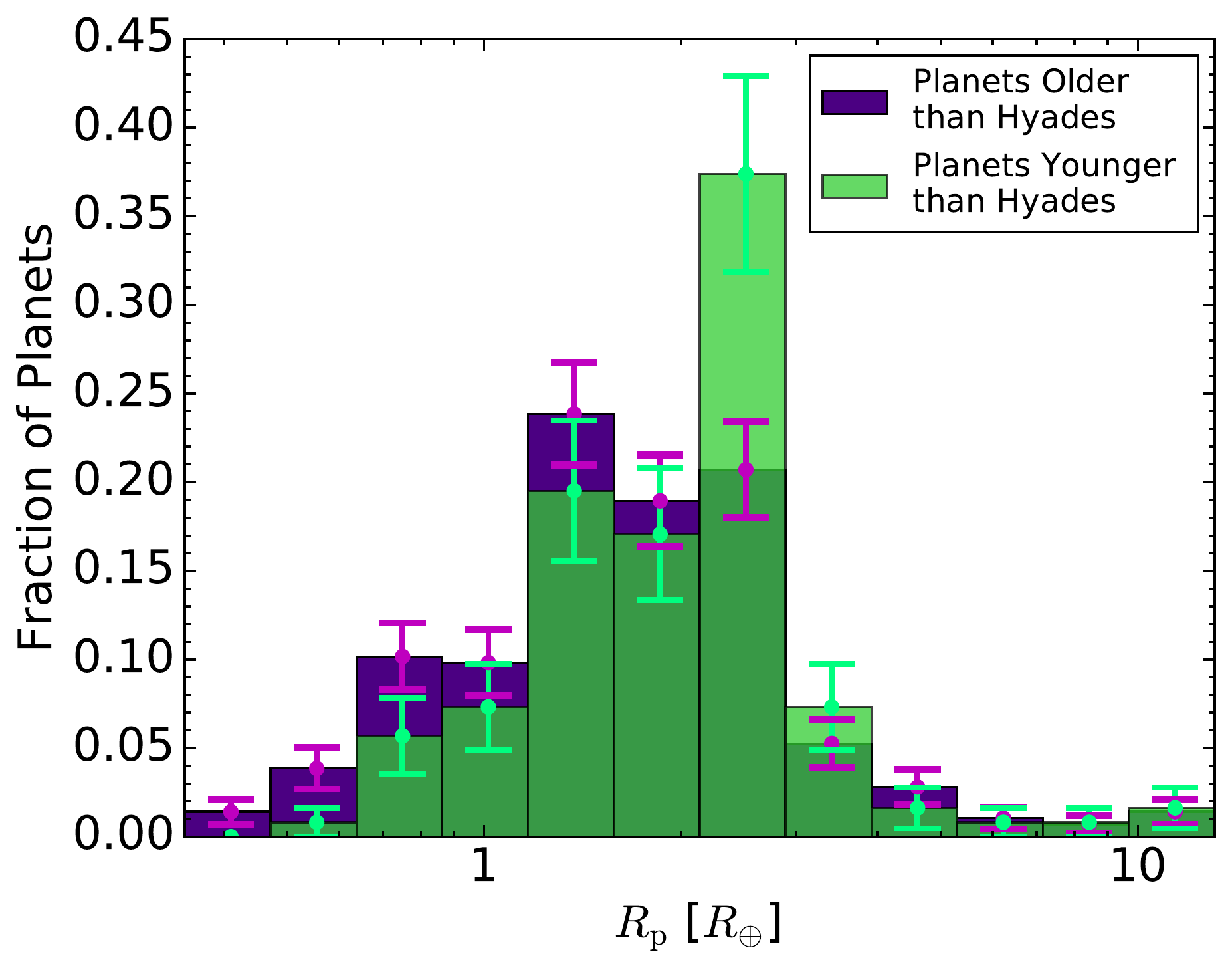}
\includegraphics[width=3.35in]{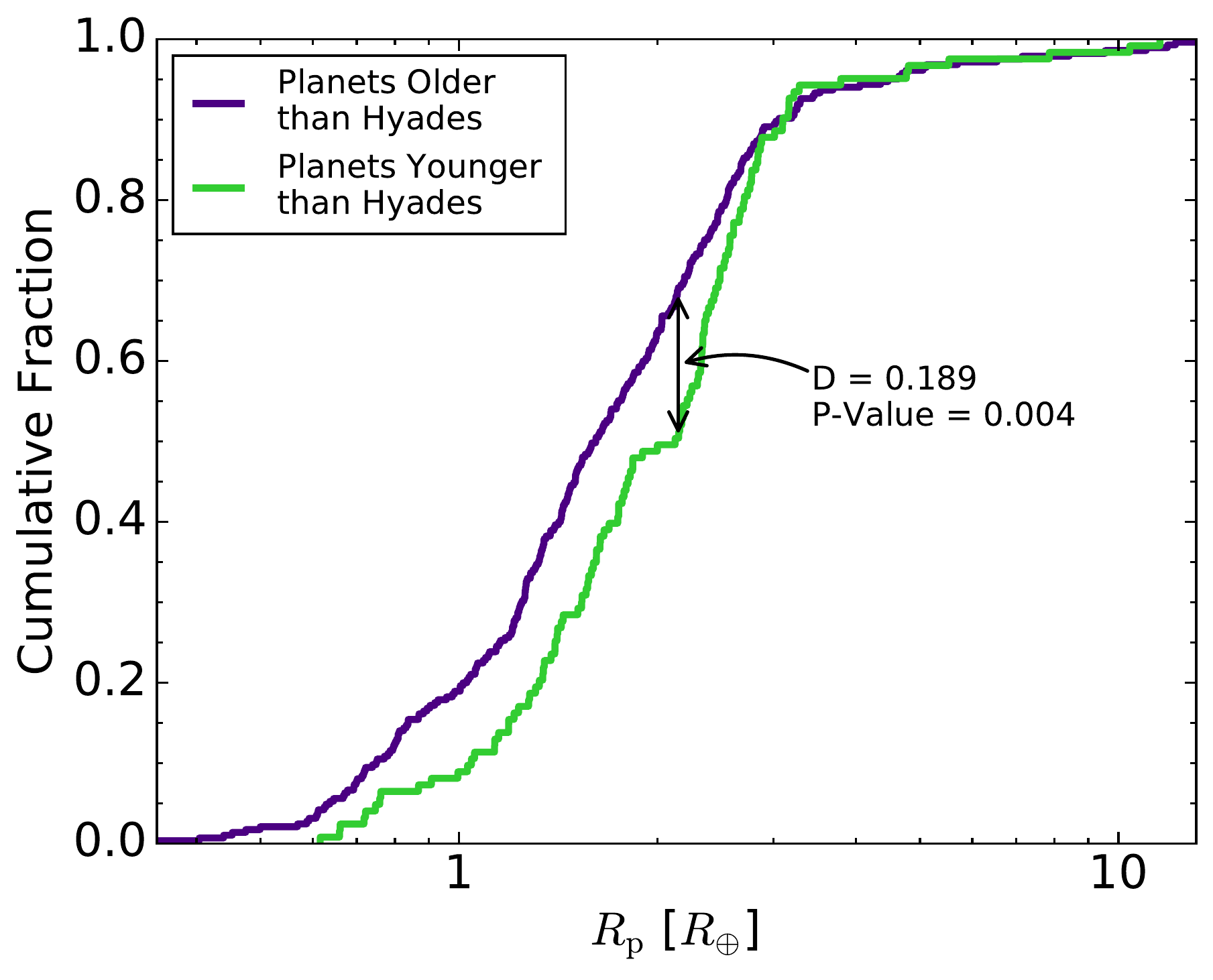}
\caption{Planet radius distributions for planets older (purple) and younger (green) than the Hyades, while all planets meet the following criteria:  host spectral $S/N$ $>$ 10, host \teff\ $<$ 5500 K, host is on the main sequence, planet is not a false positive, planet impact parameter $b$ $<$ 0.7, and planet transit $S/N$ $>$ 10. $Top$: Normalized histograms for $R_{\mathrm{p}}$. We include Poisson error bars on each of the bins for reference. $Bottom$: Cumulative fraction of planets as a function of $R_{\mathrm{p}}$. We have labeled the K-S statistic, which represents the greatest distance between the two distributions, as D. The largest difference occurs at $R_{\mathrm{p}} \simeq 2.0R_{\oplus}$. The labeled p-value indicates that, at $\sim$3$\sigma$ significance, we can reject the null hypothesis that the two samples come from the same parent distribution.} \label{fig:oldyoungcomp}
\end{figure}

\subsection{A Statistical Comparison of Old and Young Systems}
To differentiate between young and old systems, we use the Hyades as the dividing line. The Hyades is $\sim$650 Myr old \citep{Perryman1998,Brandt2015,Boesgaard2016}; systems that fall above (below) it in A(Li) versus \teff\ space (see Figure \ref{fig:aliteffpradcomp}) are younger (older) than the Hyades. This statement holds true for the majority of systems, but those close to the Hyades curve are more likely to be on the wrong side of the Hyades curve. Errors in measurement and potential abundance effects ($i.e.$ differences in initial A(Li)) affect our young/old system designation in detail. We do not expect the measurement errors to systematically bias our results. In addition, while abundance effects may introduce a systematic bias ($i.e.$ Kepler stars have systematically higher initial A(Li) compared to the Hyades), ensembles of stars have been shown to give consistent initial A(Li) although the individual scatter may be large \citep{Soderblom2010}. Much like the measurement errors are unlikely to introduce systematic bias, scatter in the initial A(Li) for our Kepler stars should not introduce systematic effects in our reported A(Li). We note that the Hyades's curve does not extend far enough at low \teff\ for us to compare systems with stellar \teff\ $\lesssim 5100$ K. Therefore, we extrapolate the Hyades curve to low \teff\ to solve this issue. We choose the observed median curve from the CKS sample as our adopted extrapolation (blue dashed line at low \teff\ in A(Li)--\teff\ plots) because it appears to follow the A(Li)-\teff\ relationship of the Hyades at low \teff.

Before we begin any detailed statistical tests, we must again eliminate any systems which may introduce bias. We removed all planets/systems discussed at the beginning of \S4.4, in addition to systems with \teff\ $>$ 5500 K because our sample of young systems at higher \teff\ is incomplete. However, we do re-include stars with spectral $S/N$ = 10--30 because some of these systems include significant Li detections. We also note that ignoring these stars significantly reduces the size of our available sample. After making these cuts, 257 systems that host 408 exoplanet candidates remained.

\begin{figure}[t]
\includegraphics[width=3.35in]{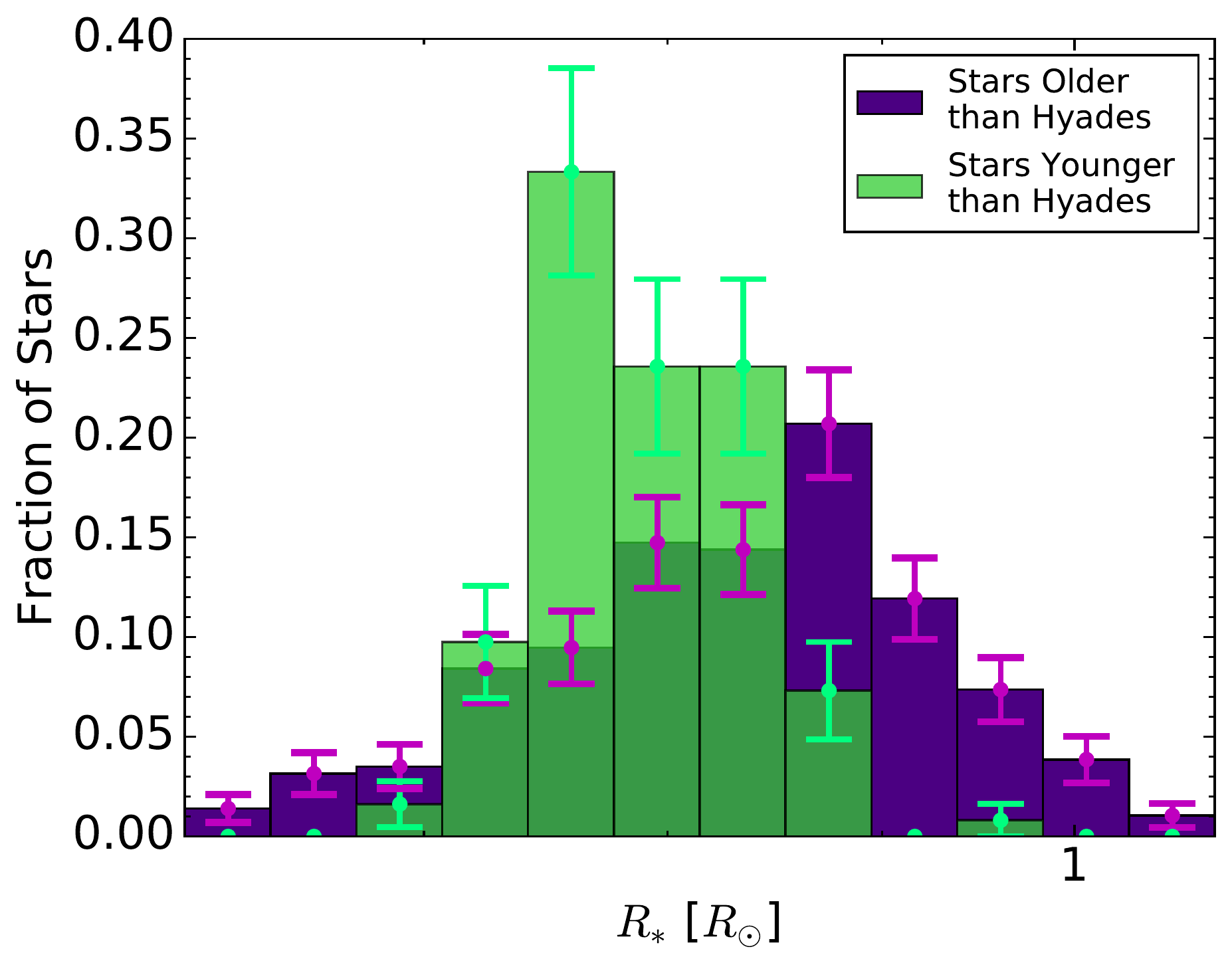}
\caption{Stellar radius normalized histograms for stars older (purple) and younger (green) than the Hyades, while all stars meet the following criteria:  spectral $S/N$ $>$ 10, \teff\ $<$ 5500 K, star is on the main sequence, hosted planet is not a false positive, hosted planet impact parameter $b$ $<$ 0.7, and hosted planet transit $S/N$ $>$ 10. We include Poisson error bars on each of the bins for reference.} \label{fig:oldyoungsrad}
\end{figure}

We separated the two groups into old ($\Delta_{\mathrm{A(Li),Hyades}} \leq 0$) and young ($\Delta_{\mathrm{A(Li),Hyades}} > 0$) systems. We placed these points into $R_{\mathrm{p}}$ bins and plotted the resulting normalized histograms (see top plot of Figure \ref{fig:oldyoungcomp}). There are 285 old planets (purple histogram) and 123 young planets (green histogram). We also performed a two-sided/two-sample K-S test to determine if the two distributions are from different parent populations. We plot the cumulative fraction of planets in $R_{\mathrm{p}}$ and the K-S test result in the bottom panel of Figure \ref{fig:oldyoungcomp}. With a p-value of 0.004, we reject the hypothesis that the two samples were drawn from the same parent distribution at a statistical significance of $\sim$3$\sigma$. Therefore, we conclude that old and young systems represent distinct populations in \rp\ space. The median value for the young planets is \rp\ = 2.13 $\pm$ 0.01 \rearth, while the median value for the old planets is \rp\ = 1.61 $\pm$ 0.01 \rearth.

To ensure we are not being fooled by potential confounding factors (such as transit $S/N$ limitations of small planets around larger stars), we plot the stellar radius distributions \citep{Petigura2017} of stars younger (green) and older (purple) than the Hyades in Figure \ref{fig:oldyoungsrad} as we do planets in Figure \ref{fig:oldyoungcomp}. From Figure \ref{fig:oldyoungsrad}, we observe that the stars younger than the Hyades are systematically smaller than their older counterparts. This limits the possibilities for artificial inflation of younger planets. We note that the stars are primarily in the range of 0.6--1.0 $R_{\mathrm{\odot}}$. From transit $S/N$ considerations and these stellar radius distributions, it is likely that we can detect smaller planets around the younger stars if they were present. Ultimately, we conclude that our young stars are smaller and unlikely to explain the difference we see in the old and young planet radii distributions.

\begin{figure}[t]
\includegraphics[width=3.35in]{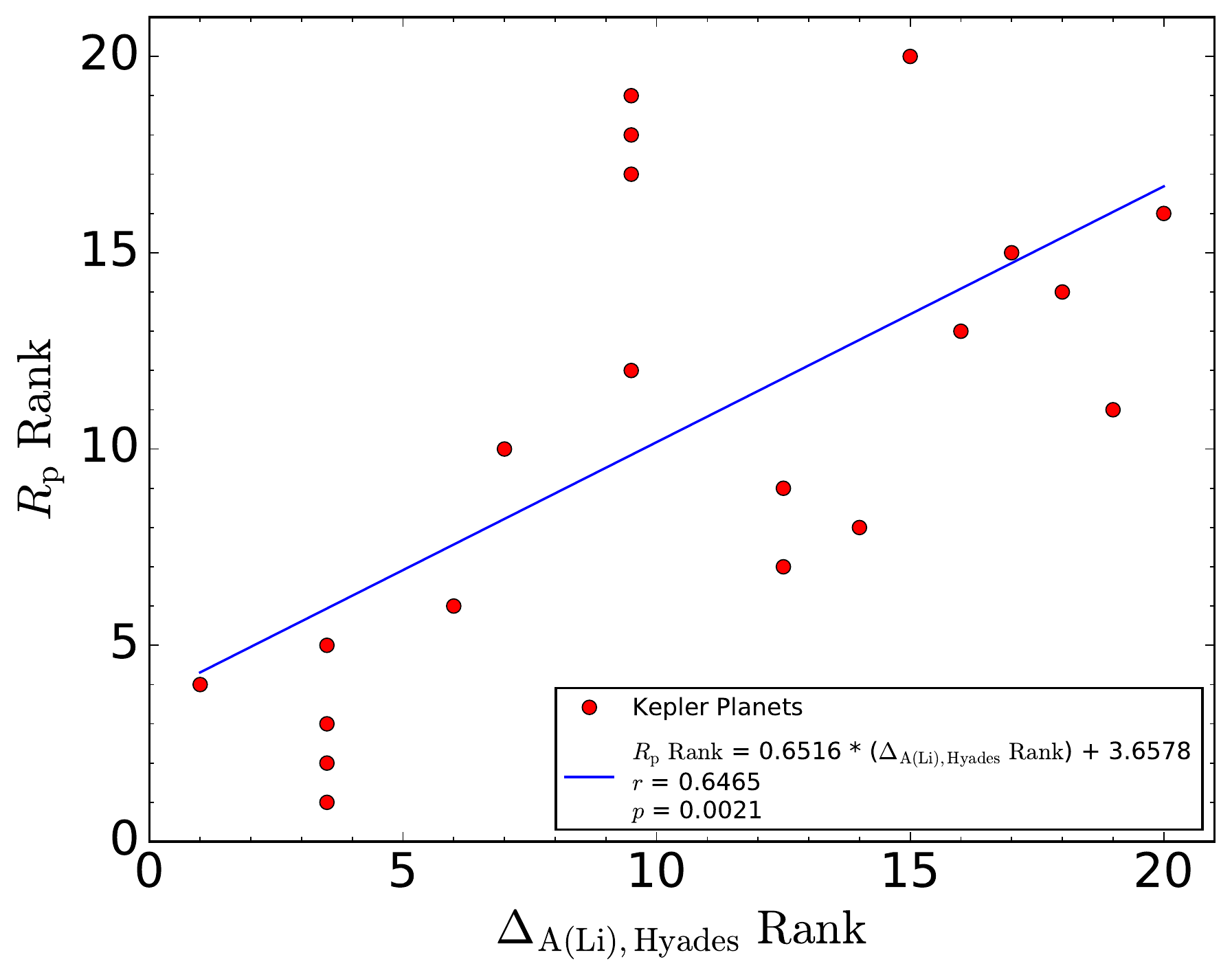}
\caption{Spearman rank-order correlation coefficient test for $R_{\mathrm{p}}$ and $\Delta_{\mathrm{A(Li),Hyades}}$. We have removed all subgiant and giant stars, stars with \teff\ $>$ 5500 K, stars with upper limit A(Li), stars with $\Delta_{\mathrm{A(Li),Hyades}} < 0$ (old stars), false positive planets, planets with $b > 0.7$, and planets with transit $S/N$ $<$ 10. The red points are the rank-orders of the remaining Kepler planets in $R_{\mathrm{p}}$ and $\Delta_{\mathrm{A(Li),Hyades}}$, while the blue line is the line of best fit. The legend includes the equation for the line of best fit, as well as $r$, the correlation coefficient, and $p$, the likelihood that our two parameters are uncorrelated.} \label{fig:corrtest}
\end{figure}

As has been demonstrated by \cite{Babu2006}, the one sample K-S test and other empirical distribution functions can be unreliable when using them to determine the parameters of best-fit models. Fortunately, we are using a two sample K-S test to compare two populations within our data -- not using it to determine parameters of those populations. Despite the utility of the two sample K-S statistic, according to \cite{Engmann2011}, this statistic is inferior to the two sample Anderson-Darling (A-D) statistic. The A-D test is more proficient in detecting differences in shift, scale, and symmetry between samples from two different distributions. Thus, we perform a two sample A-D test on the same sample used for the two sample K-S test. Our resulting normalized A-D statistic is 5.98, corresponding to a p-value of 0.002 ($>3\sigma$). Unsurprisingly, we report a more significant A-D p-value than K-S p-value.

However, these results are not robust:  if we remove systems with $S/N$ $<$ 30 from our sample, our K-S p-value rises to 0.48 and our A-D p-value rises to 0.37, which both correspond to a statistical significance of $<$ 1$\sigma$. This is not sufficiently significant to definitively conclude that exoplanet radii decrease as exoplanets age. Removing these moderate-$S/N$ ($10 < S/N < 30$) systems reduces the number of old planets from 285 to 212 and the number of young planets from 123 to 43. We argue that including moderate-$S/N$ systems is important for the K-S and A-D tests because without them we are left with a very small sample of younger planets, insufficient for statistical distribution comparisons with the older planets.

We also perform additional statistical analyses, including Pearson and Spearman rank-order correlation coefficient tests of the relation between \rp\ and $\Delta_{\mathrm{A(Li),Hyades}}$. We use these tests to analyze 20 exoplanet candidates that remain after removing all of the following stars and planets:  subgiants and giants, \teff\ $>$ 5500 K, upper limit A(Li), stars with $\Delta_{\mathrm{A(Li),Hyades}} < 0$ (old stars), stars with spectral $S/N$ $<$ 10, false positive planets, planets with $b > 0.7$, and planets with transit $S/N$ $<$ 10. We remove all upper limits in A(Li) because ages of these systems are inherently uncertain. In addition, we eliminate old stars and planets because these systems/planets contaminate the theoretically predicted relationship between planet size and age in younger systems \citep{Lopez2012}. The Spearman rank-order test is more reliable than the Pearson test because it assumes nothing about the underlying relationship between planet radius and age, unlike the Pearson test which assumes a linear relationship.

We plot the results of the Spearman test in Figure \ref{fig:corrtest}, and we report that the Spearman test returns a positive correlation between planet radius and A(Li) relative to the Hyades. The figure has a strong, positive correlation coefficient ($r$ = 0.6465), and a small p-value ($p$ = 0.0021). We performed a similar analysis using the Pearson correlation coefficient test for comparison. Unsurprisingly, the Spearman correlation coefficient is stronger than the Pearson coefficient. Unlike the K-S and A-D test results, these correlation coefficients are robust and not sensitive to changes in parameter ranges. Although the p-value indicates a high significance of correlation between planet radius and the relative age of these systems, our small sample size makes it difficult to trust these p-values at their reported significance. Nevertheless, the combination of the low p-value and the strong, positive correlation coefficient between planet radius and age support the current models of young exoplanet evolution.

\section{Discussion}
Our findings are suggestive that larger planets are more likely to orbit younger stars. \cite{Lopez2012} discuss a few possible mechanisms, including cooling and atmosphere loss, that cause planets to contract over time. The authors choose typical timescales of 10 Myr and 100 Myr for cooling and contraction after formation. Notably, the youngest stars in our sample fall close to the 100 Myr isochrone from \cite{Xiong2009}, which may indicate that these planets are still inflated from a combination of residual heat from formation and extreme ultraviolet (XUV) flux received from their host star. Interestingly, the planet radius gap discussed in \cite{Fulton2017} separates the calculated median planet radius for older planets (\rp\ = 1.61 $R_{\oplus}$) from the calculated median planet radius for younger planets (\rp\ = 2.13 $R_{\oplus}$). 

Additionally, \cite{Lopez2012} investigate the $Kepler$-11 system and the possible formation mechanisms for each of the planetary companions. Figure 2 in this paper provides an illustration of the degeneracy in producing the radius of $Kepler$-11b. Its current radius could be just as easily explained by a 0.3\% H/He composition or an 11\% H/He composition. The authors state that most of the mass (and radius) loss happens in the first Gyr. Our finding of larger planets orbiting younger stars agree with their models.

Heat from formation and XUV radiation are not the only parameters that impact mass (and radius) loss. \cite{Lopez2013}, in looking at the $Kepler$-36 system, found that core mass plays a large role in the evolution of the radius of a planet. However, $Kepler$ data do not include the masses of most of the CKS planets, so it is difficult for us to conclude anything about their compositions and whether that has an effect on the observed distribution of young systems.

Moreover, \cite{Lopez2012} considered the hydrodynamic mass loss of close-in, low-mass, low-density (LMLD) planets from the XUV radiation released by their young MS stars. Mass loss rates are much larger when planets are young because of (1) planetary radii are considerably larger due to heat from formation and (2) a star's $F_{\mathrm{XUV}}$ is $\sim$500 times higher at 100 Myr than the same star's $F_{\mathrm{XUV}}$ at a few Gyrs \citep{Lopez2012}. \cite{Fortney2007} provides a plot of \rp\ versus age for a few masses ranging from 0.1--3.0 $M_{\mathrm{J}}$ at various orbital radii. Planets closer than 0.045 AU to a solar analog star are significantly affected by the XUV radiation, while those further away are minimally impacted. Therefore, we calculated the corresponding insolation flux at 0.045 AU ($\approx$500 $F_{\oplus}$), and used this criterion to limit our sample of young and old planets. All planets with $F_{\mathrm{p}} > 500 F_{\oplus}$ were excluded to determine if, statistically, we could separate residual heat from formation inflation from XUV heating inflation. However, we found the old and young planet \rp\ distributions were more similar than those with no exclusions based on $F_{\mathrm{p}}$.

Because we ignored evolved stars and the number of Jupiter-size planets included in the CKS is small, we were unable to test the planet reinflation theories detailed in \cite{Grunblatt2016}. Nevertheless, our results do provide evidence for the shrinking of planets as they orbit their main sequence stars; interestingly, planet shrinkage during their host's main sequence lifetime is an initial condition for post-main sequence planet reinflation.

We find no evidence for a correlation between between age and other planet parameters, such as $Kepler$ planet multiplicity, orbital period, and insolation flux. This is unsurprising on the large scales we consider in this paper. $Kepler$ planet multiplicity is inherently uncertain given that our planet detections for any star are, by no means, complete. We may expect to see less planets around the older stars due to dynamical interactions (and potential planet ejection) in those systems, but these interactions typically happen very early in the host's lifetime. We expect the incompleteness of our planet sample to trump any age effects. Orbital periods are extremely precise compared to other planet property determinations, but it is unclear whether there exist any processes to systematically bias old planets' orbital periods relative to young planets' orbital periods. The variation in orbital period from system to system likely dominates over proposed processes such as planet migration that can lead to different old and young planet populations. In addition, insolation flux is mostly a function of stellar luminosity ($i.e.$ stellar mass/radius), which, as Figure \ref{fig:oldyoungsrad} illustrates, introduces systematic bias between systems older and younger than the Hyades. Thus, any age effects on insolation flux are insignificant compared to the differences in initial conditions.

\section{Summary and Conclusions}
In this paper, we have detailed our automated Li pipeline from normalization of the spectra to the determination of A(Li). With these data, we produced a catalog of A(Li), which includes the relevant stellar properties and errors. We proceeded to compare stellar properties (\logg\ and [Fe/H]) with A(Li) through A(Li)-\teff\ plots, and found no trends with A(Li). Additionally, we compared exoplanet properties using the same A(Li)-\teff\ plots. We found that most exoplanet properties ($Kepler$ planet disposition, multiplicity, orbital period, and insolation flux) do not trend with A(Li). Raw A(Li) values are not the best age differentiator, as A(Li) varies with \teff, so we use our interpolated empirical Hyades curve (Figure \ref{fig:hyades}) derived from \cite{Boesgaard2016} to separate systems older and younger than 650 Myr. Because A(Li) relative to the Hyades is a proxy for the age of FGK main sequence dwarfs, we conclude that these exoplanet properties show no trends with relative age. However, we do find statistical evidence for the shrinking of exoplanet radii with age based on K-S, Pearson, and Spearman tests.

We conclude that the difference in the \rp\ distributions of young and old systems suggests exoplanet radii shrink as they age during their host's main sequence lifetime, a phenomenon that may result from a combination of photoevaporation of the planets' atmospheres and cooling (and contraction) from the planets' residual heat from formation. We look forward to future surveys that link exoplanet properties and age. These studies may reveal paramount information about the mechanisms of exoplanet formation and evolution, as well as the processes behind our own Solar System's origin.

\acknowledgements
T.A.B. thanks Daniel Huber, Sam Grunblatt, Conor McPartland, Benjamin Boe, Christian Flores, Maissa Salama, and Zhoujian Zhang for useful discussion regarding the pipeline and statistical tests. T.A.B. also thanks Michael Lum for his model atmosphere interpolation code and guidance with MOOG and Christopher Waters for his assistance with equivalent width fitting packages. 

A.\,W.\,H.\ acknowledges NASA grant NNX12AJ23G. This research has made use of the NASA Exoplanet Archive, which is operated by the California Institute of Technology, under contract with the National Aeronautics and Space Administration under the Exoplanet Exploration Program. 

The authors thank Joshua Winn, Benjamin Fulton, David Soderblom, and Howard Isaacson for feedback and discussion on previous drafts of this paper. The authors also thank the many observers who helped gather these spectra. Additionally, the authors thank the reviewer for the useful comments on the submitted drafts of this paper. The data presented herein were obtained at the W.\ M.\ Keck Observatory, which is operated as a scientific partnership among the California Institute of Technology, the University of California and the National Aeronautics and Space Administration. The Observatory was made possible by the generous financial support of the W.\ M.\ Keck Foundation.  The authors thank the time allocation committees of the University of Hawaii, the University of California, and the California Institute of Technology for large allocations of telescope time. Finally, the authors wish to recognize and acknowledge the very significant cultural role and reverence that the summit of Mauna Kea has always had within the indigenous Hawaiian community. We are most fortunate to have the opportunity to conduct observations from this mountain.

\software{MOOG \citep{Sneden2012}, LMFIT \citep{Newville2014}, PyRAF \citep{Pyraf2012}, SciPy \citep{Scipy}, Pandas \citep{Pandas}, Matplotlib \citep{Hunter2007}}

\begin{deluxetable*}{lccccccrcrr}
\tabletypesize{\scriptsize}
\tablenum{1}
\tablewidth{0pt}
\tablecolumns{8}
\tablecaption{Pipeline Results for $Kepler$ Planet Host Stars}
\tablehead{
\colhead{Obs Code} & \colhead{Obs Date} & \colhead{KOI} & \colhead{KepMag} & \colhead{$S/N$} & \colhead{\teff\ [K]} & \colhead{\logg\ [dex]} & \colhead{[Fe/H]} & \colhead{$\xi$ [km/s]} & \colhead{$EW_{\mathrm{Li}}$ [m\AA]} & \colhead{A(Li) [dex]}}
\def\arraystretch{1.0}
\startdata
j122.742 & 2011-06-16 & 1 & 11.34 & 39 & 5819 & 4.40 & 0.01 & 1.04 & 85.5 $\pm$ 7.9 & 2.62 $\pm$ 0.08\\
j122.92 & 2011-06-13 & 2 & 10.46 & 39 & 6449 & 4.13 & 0.20 & 1.77 & 82.9 $\pm$ 7.3 & 3.11 $\pm$ 0.07\\
j122.81 & 2011-06-13 & 3 & 9.17 & 41 & 4864 & 4.50 & 0.33 & 0.54 & 4.2 $\pm$ 5.1 & $<$ -0.40\\
j70.1247 & 2009-06-05 & 6 & 12.16 & 119 & 6348 & 4.36 & 0.04 & 1.58 & 15.4 $\pm$ 2.2 & 2.16 $\pm$ 0.08\\
j74.509 & 2009-07-31 & 7 & 12.21 & 126 & 5827 & 4.09 & 0.18 & 1.17 & 54.0 $\pm$ 2.1 & 2.36 $\pm$ 0.06\\
j70.1251 & 2009-06-05 & 8 & 12.45 & 89 & 5891 & 4.54 & -0.07 & 1.05 & 54.5 $\pm$ 3.6 & 2.42 $\pm$ 0.06\\
j77.875 & 2009-10-05 & 10 & 13.56 & 74 & 6181 & 4.24 & -0.08 & 1.46 & 19.7 $\pm$ 3.6 & 2.15 $\pm$ 0.10\\
j72.483 & 2009-07-03 & 17 & 13.30 & 77 & 5660 & 4.15 & 0.36 & 1.09 & 2.3 $\pm$ 2.3 & $<$ 0.75\\
j72.487 & 2009-07-03 & 18 & 13.37 & 73 & 6332 & 4.12 & 0.02 & 1.66 & 52.3 $\pm$ 3.8 & 2.75 $\pm$ 0.06\\
j93.303 & 2010-06-26 & 20 & 13.44 & 103 & 5927 & 4.01 & 0.03 & 1.30 & 40.7 $\pm$ 2.6 & 2.31 $\pm$ 0.06\\
j76.1283 & 2009-09-05 & 22 & 13.44 & 79 & 5891 & 4.21 & 0.21 & 1.19 & 25.8 $\pm$ 3.7 & 2.04 $\pm$ 0.09\\
j126.89 & 2011-07-09 & 41 & 11.20 & 207 & 5854 & 4.07 & 0.10 & 1.21 & 34.3 $\pm$ 1.3 & 2.15 $\pm$ 0.06\\
j90.90 & 2010-05-01 & 42 & 9.36 & 212 & 6306 & 4.28 & -0.01 & 1.57 & 15.7 $\pm$ 1.3 & 2.14 $\pm$ 0.06\\
j124.500 & 2011-06-23 & 46 & 13.77 & 50 & 5661 & 4.07 & 0.39 & 1.12 & 98.2 $\pm$ 5.5 & 2.57 $\pm$ 0.07\\
j120.1133 & 2011-05-26 & 49 & 13.70 & 46 & 5779 & 4.34 & -0.06 & 1.10 & 3.8 $\pm$ 4.6 & $<$ 1.07\\
j130.1072 & 2011-08-18 & 63 & 11.58 & 181 & 5673 & 4.68 & 0.25 & 0.94 & 90.3 $\pm$ 1.6 & 2.41 $\pm$ 0.06\\
j76.1081 & 2009-09-04 & 64 & 13.14 & 92 & 5357 & 3.86 & 0.09 & 1.01 & 3.3 $\pm$ 2.9 & $<$ 0.62\\
j76.1276 & 2009-09-05 & 69 & 9.93 & 163 & 5594 & 4.41 & -0.09 & 0.97 & 1.2 $\pm$ 1.6 & $<$ 0.38\\
j97.1478 & 2010-08-24 & 70 & 12.50 & 133 & 5508 & 4.47 & 0.11 & 0.91 & 3.3 $\pm$ 2.1 & $<$ 0.71\\
\nodata & \nodata & \nodata & \nodata & \nodata & \nodata & \nodata & \nodata & \nodata & \nodata & \nodata\\
\enddata
\tablecomments{Observational and stellar data for the CKS stars analyzed in this paper. The stellar parameters \teff, \logg, and [Fe/H] have uncertainties of 60 K, 0.10 dex, and 0.04 dex, respectively; these are the adopted values from \cite{Petigura2017}. The KepMag column is from the NASA Exoplanet Archive. The other columns ($S/N$, $\xi$, $EW_{\mathrm{Li}}$, and A(Li)) and their uncertainties, where relevant, were computed by our pipeline. We employ \cite{Takeda2013}'s treatment of $\xi$. All items in the table with $<$ symbols in the final column indicate stars with $EW_{\mathrm{Li}}$ $<$ $\sigma_{\mathrm{UL}} + \sigmaeweq$, which we have flagged as upper limits in our analysis (and downward arrows in our plots). The full table, in machine-readable format, can be found online.} \label{tab:full}
\end{deluxetable*}

\begin{deluxetable*}{lccccccrr}
\tabletypesize{\scriptsize}
\tablenum{2}
\tablewidth{0pt}
\tablecolumns{9}
\tablecaption{Stars Younger than the Hyades}
\tablehead{
\colhead{KOI} & \colhead{$S/N$} & \colhead{\teff\ [K]} & \colhead{\logg\ [dex]} & \colhead{[Fe/H]} & \colhead{$\xi$ [km/s]} & \colhead{$EW_{\mathrm{Li}}$ [m\AA]} & \colhead{A(Li) [dex]} & \colhead{$\Delta_{\mathrm{A(Li),Hyades}}$ [dex]}}
\def\arraystretch{1.0}
\startdata
1 & 39 & 5819 & 4.40 & 0.01 & 1.04 & 85.5 $\pm$ 7.9 & 2.62 $\pm$ 0.08 & 0.08\\
2 & 39 & 6449 & 4.13 & 0.20 & 1.77 & 82.9 $\pm$ 7.3 & 3.11 $\pm$ 0.07 & 0.59\\
46 & 50 & 5661 & 4.07 & 0.39 & 1.12 & 98.2 $\pm$ 5.5 & 2.57 $\pm$ 0.07 & 0.47\\
63 & 181 & 5673 & 4.68 & 0.25 & 0.94 & 90.3 $\pm$ 1.6 & 2.41 $\pm$ 0.06 & 0.27\\
98 & 71 & 6500 & 4.22 & 0.05 & 1.78 & 52.1 $\pm$ 4.0 & 2.87 $\pm$ 0.06 & 0.62\\
119 & 47 & 5681 & 3.86 & 0.37 & 1.19 & 100.6 $\pm$ 6.4 & 2.60 $\pm$ 0.07 & 0.43\\
149 & 42 & 5708 & 3.95 & 0.03 & 1.18 & 55.3 $\pm$ 7.2 & 2.27 $\pm$ 0.09 & 0.02\\
323 & 53 & 5529 & 4.72 & 0.10 & 0.84 & 52.9 $\pm$ 5.4 & 1.98 $\pm$ 0.09 & 0.42\\
331 & 42 & 5506 & 3.90 & 0.11 & 1.08 & 40.2 $\pm$ 6.8 & 1.91 $\pm$ 0.11 & 0.44\\
660 & 53 & 5320 & 3.89 & -0.07 & 0.98 & 25.6 $\pm$ 5.4 & 1.49 $\pm$ 0.13 & 0.61\\
684 & 59 & 5287 & 3.84 & 0.10 & 0.98 & 26.3 $\pm$ 5.0 & 1.47 $\pm$ 0.12 & 0.67\\
720 & 50 & 5260 & 4.68 & 0.04 & 0.71 & 14.5 $\pm$ 5.5 & 1.11 $\pm$ 0.23 & 0.39\\
853 & 28 & 4876 & 4.73 & -0.02 & 0.47 & 43.8 $\pm$ 11.0 & 1.17 $\pm$ 0.20 & 1.18\\
1019 & 156 & 5018 & 3.55 & 0.17 & 0.92 & 7.1 $\pm$ 1.7 & 0.51 $\pm$ 0.14 & 0.32\\
1117 & 48 & 6513 & 4.16 & -0.02 & 1.82 & 38.3 $\pm$ 6.4 & 2.72 $\pm$ 0.10 & 0.53\\
1175 & 43 & 5640 & 3.80 & 0.10 & 1.19 & 46.8 $\pm$ 7.1 & 2.12 $\pm$ 0.10 & 0.08\\
1199 & 39 & 4772 & 4.53 & 0.11 & 0.48 & 85.9 $\pm$ 8.7 & 1.28 $\pm$ 0.13 & 1.40\\
1208 & 46 & 6417 & 4.16 & -0.07 & 1.73 & 48.3 $\pm$ 6.1 & 2.77 $\pm$ 0.08 & 0.12\\
1221 & 106 & 5002 & 3.62 & 0.33 & 0.89 & 9.0 $\pm$ 2.4 & 0.52 $\pm$ 0.16 & 0.36\\
1230 & 51 & 5119 & 3.35 & 0.01 & 1.04 & 26.8 $\pm$ 5.4 & 1.31 $\pm$ 0.13 & 0.92\\
1413 & 34 & 5253 & 3.79 & -0.03 & 0.98 & 18.1 $\pm$ 8.3 & 1.26 $\pm$ 0.29 & 0.56\\
1438 & 33 & 5722 & 3.94 & 0.19 & 1.19 & 64.2 $\pm$ 9.5 & 2.36 $\pm$ 0.10 & 0.07\\
1463 & 136 & 6532 & 4.19 & -0.04 & 1.83 & 38.2 $\pm$ 2.0 & 2.73 $\pm$ 0.05 & 0.66\\
1800 & 43 & 5621 & 4.69 & 0.07 & 0.90 & 100.1 $\pm$ 6.4 & 2.49 $\pm$ 0.08 & 0.51\\
1839 & 51 & 5517 & 4.67 & 0.17 & 0.85 & 99.4 $\pm$ 5.3 & 2.28 $\pm$ 0.08 & 0.77\\
1864 & 42 & 5620 & 3.93 & 0.12 & 1.14 & 52.1 $\pm$ 7.7 & 2.15 $\pm$ 0.10 & 0.18\\
1985 & 42 & 4950 & 4.64 & 0.01 & 0.54 & 10.3 $\pm$ 5.9 & 0.62 $\pm$ 0.40 & 0.55\\
2033 & 38 & 5051 & 4.53 & -0.12 & 0.64 & 12.9 $\pm$ 7.5 & 0.61 $\pm$ 0.40 & 0.35\\
2035 & 48 & 5558 & 4.67 & 0.10 & 0.87 & 56.2 $\pm$ 5.5 & 2.06 $\pm$ 0.09 & 0.38\\
2046 & 45 & 5579 & 4.07 & 0.23 & 1.07 & 44.4 $\pm$ 6.4 & 2.04 $\pm$ 0.10 & 0.25\\
2115 & 13 & 5239 & 4.71 & 0.13 & 0.68 & 95.3 $\pm$ 23.1 & 1.90 $\pm$ 0.20 & 1.24\\
2175 & 42 & 5459 & 3.91 & 0.13 & 1.06 & 43.8 $\pm$ 7.1 & 1.91 $\pm$ 0.11 & 0.61\\
2228 & 41 & 6656 & 4.19 & -0.10 & 1.95 & 47.0 $\pm$ 6.7 & 2.92 $\pm$ 0.09 & 4.38\\
2261 & 36 & 5176 & 4.70 & 0.06 & 0.65 & 31.4 $\pm$ 7.7 & 1.34 $\pm$ 0.17 & 0.83\\
2479 & 43 & 5372 & 3.88 & 0.08 & 1.02 & 28.7 $\pm$ 6.7 & 1.61 $\pm$ 0.14 & 0.59\\
2516 & 35 & 5431 & 3.91 & 0.30 & 1.04 & 19.1 $\pm$ 8.1 & 1.48 $\pm$ 0.26 & 0.27\\
2541 & 40 & 5090 & 3.70 & 0.15 & 0.91 & 15.9 $\pm$ 6.9 & 0.96 $\pm$ 0.27 & 0.62\\
2639 & 24 & 5583 & 3.89 & -0.02 & 1.13 & 99.4 $\pm$ 13.2 & 2.50 $\pm$ 0.11 & 0.69\\
2640 & 48 & 4896 & 3.02 & -0.13 & 1.01 & 10.6 $\pm$ 5.6 & 0.63 $\pm$ 0.34 & 0.61\\
2675 & 45 & 5756 & 4.63 & 0.13 & 1.00 & 82.1 $\pm$ 6.2 & 2.53 $\pm$ 0.07 & 0.13\\
2678 & 43 & 5416 & 4.70 & 0.12 & 0.78 & 136.4 $\pm$ 6.4 & 2.42 $\pm$ 0.09 & 1.26\\
2748 & 44 & 5499 & 4.00 & -0.06 & 1.05 & 33.0 $\pm$ 6.2 & 1.81 $\pm$ 0.12 & 0.37\\
2769 & 54 & 5787 & 3.96 & -0.02 & 1.22 & 69.4 $\pm$ 5.8 & 2.47 $\pm$ 0.07 & 0.00\\
2831 & 38 & 5752 & 3.92 & -0.01 & 1.21 & 65.8 $\pm$ 7.1 & 2.40 $\pm$ 0.09 & 0.02\\
2859 & 46 & 5260 & 4.51 & -0.07 & 0.76 & 12.4 $\pm$ 6.1 & 0.98 $\pm$ 0.32 & 0.26\\
2885 & 45 & 5492 & 3.93 & -0.36 & 1.07 & 27.5 $\pm$ 6.1 & 1.70 $\pm$ 0.13 & 0.29\\
2891 & 47 & 6142 & 4.02 & 0.14 & 1.51 & 99.8 $\pm$ 5.8 & 2.99 $\pm$ 0.06 & 0.01\\
3012 & 39 & 5493 & 4.16 & -0.50 & 1.00 & 24.0 $\pm$ 7.3 & 1.63 $\pm$ 0.18 & 0.22\\
3202 & 41 & 5262 & 3.80 & 0.03 & 0.98 & 23.9 $\pm$ 6.7 & 1.40 $\pm$ 0.17 & 0.68\\
3239 & 47 & 5668 & 4.66 & 0.09 & 0.94 & 124.5 $\pm$ 5.7 & 2.70 $\pm$ 0.07 & 0.58\\
3244 & 58 & 4970 & 3.13 & -0.03 & 1.02 & 25.1 $\pm$ 4.8 & 1.11 $\pm$ 0.12 & 1.01\\
3371 & 45 & 5428 & 4.56 & 0.00 & 0.84 & 54.1 $\pm$ 7.2 & 1.98 $\pm$ 0.10 & 0.78\\
3473 & 37 & 5157 & 4.65 & 0.05 & 0.66 & 82.6 $\pm$ 7.4 & 1.85 $\pm$ 0.11 & 1.38\\
3835 & 45 & 5013 & 4.70 & 0.04 & 0.56 & 118.8 $\pm$ 6.1 & 1.92 $\pm$ 0.11 & 1.73\\
3871 & 49 & 5193 & 4.62 & 0.07 & 0.69 & 11.0 $\pm$ 5.6 & 0.87 $\pm$ 0.33 & 0.32\\
3876 & 40 & 5720 & 4.64 & 0.12 & 0.98 & 119.8 $\pm$ 6.7 & 2.73 $\pm$ 0.07 & 0.45\\
3886 & 47 & 4760 & 2.94 & 0.20 & 0.96 & 11.5 $\pm$ 6.3 & 0.45 $\pm$ 0.36 & 0.58\\
3891 & 52 & 5080 & 3.75 & -0.10 & 0.89 & 21.6 $\pm$ 5.2 & 1.08 $\pm$ 0.15 & 0.76\\
3908 & 49 & 5721 & 3.88 & 0.03 & 1.21 & 58.1 $\pm$ 6.6 & 2.30 $\pm$ 0.09 & 0.02\\
3936 & 55 & 5081 & 4.66 & 0.17 & 0.61 & 162.5 $\pm$ 4.8 & 2.03 $\pm$ 0.12 & 1.71\\
3991 & 41 & 5606 & 4.62 & -0.02 & 0.92 & 96.7 $\pm$ 6.6 & 2.48 $\pm$ 0.08 & 0.56\\
4004 & 52 & 5739 & 4.68 & -0.05 & 0.97 & 78.6 $\pm$ 5.2 & 2.48 $\pm$ 0.07 & 0.13\\
4146 & 50 & 5092 & 4.64 & 0.24 & 0.62 & 22.4 $\pm$ 5.9 & 0.67 $\pm$ 0.19 & 0.32\\
4156 & 41 & 5807 & 4.05 & 0.24 & 1.17 & 96.8 $\pm$ 7.1 & 2.69 $\pm$ 0.07 & 0.18\\
4226 & 39 & 5844 & 3.87 & 0.35 & 1.28 & 82.6 $\pm$ 7.6 & 2.62 $\pm$ 0.08 & 0.03\\
4556 & 48 & 5437 & 3.83 & -0.11 & 1.07 & 25.5 $\pm$ 6.0 & 1.61 $\pm$ 0.14 & 0.39\\
4613 & 46 & 5443 & 4.55 & -0.13 & 0.85 & 23.6 $\pm$ 6.7 & 1.45 $\pm$ 0.17 & 0.20\\
4647 & 31 & 5166 & 3.81 & 0.23 & 0.92 & 23.2 $\pm$ 9.6 & 1.18 $\pm$ 0.25 & 0.69\\
4663 & 52 & 5545 & 3.88 & 0.30 & 1.11 & 55.1 $\pm$ 6.6 & 2.12 $\pm$ 0.09 & 0.49\\
4686 & 47 & 5698 & 3.90 & 0.33 & 1.19 & 75.6 $\pm$ 5.9 & 2.44 $\pm$ 0.07 & 0.22\\
4745 & 14 & 4781 & 4.56 & 0.02 & 0.47 & 78.8 $\pm$ 14.6 & 1.41 $\pm$ 0.16 & 1.53\\
4763 & 36 & 5695 & 3.98 & 0.17 & 1.17 & 52.4 $\pm$ 7.8 & 2.23 $\pm$ 0.10 & 0.02\\
4775 & 43 & 5210 & 3.80 & 0.47 & 0.95 & 36.6 $\pm$ 6.3 & 1.48 $\pm$ 0.11 & 0.90\\
4811 & 31 & 5572 & 3.90 & 0.05 & 1.12 & 39.9 $\pm$ 9.4 & 1.97 $\pm$ 0.15 & 0.22\\
4834 & 50 & 5030 & 3.75 & 0.40 & 0.86 & 19.5 $\pm$ 5.4 & 0.86 $\pm$ 0.16 & 0.64\\
5057 & 38 & 5004 & 3.30 & 0.00 & 0.99 & 26.5 $\pm$ 7.1 & 1.17 $\pm$ 0.16 & 1.01\\
5107 & 27 & 4933 & 3.04 & -0.13 & 1.03 & 27.4 $\pm$ 10.4 & 1.11 $\pm$ 0.24 & 1.05\\
5119 & 35 & 4984 & 3.25 & 0.08 & 0.99 & 13.7 $\pm$ 7.3 & 0.84 $\pm$ 0.35 & 0.71\\
6676 & 34 & 6493 & 4.18 & -0.19 & 1.79 & 31.1 $\pm$ 8.1 & 2.59 $\pm$ 0.15 & 0.31\\
6759 & 33 & 5494 & 3.90 & 0.36 & 1.08 & 66.2 $\pm$ 8.2 & 2.16 $\pm$ 0.09 & 0.75\\
\enddata
\tablecomments{CKS stars with Li detections younger than the Hyades. The stellar parameters \teff, \logg, and [Fe/H] have uncertainties of 60 K, 0.10 dex, and 0.04 dex, respectively; these are the adopted values from \cite{Petigura2017}. The other columns ($S/N$, $\xi$, $EW_{\mathrm{Li}}$, A(Li), and $\Delta_{\mathrm{A(Li),Hyades}}$) and their uncertainties, where relevant, were computed by our pipeline. We employ \cite{Takeda2013}'s treatment of $\xi$. We do not include upper limits in this table.} \label{tab:youngstars}
\end{deluxetable*}

\begin{deluxetable*}{lcccccrrr}
\tabletypesize{\scriptsize}
\vspace{-0.27cm}
\tablenum{3}
\tablewidth{0pt}
\tablecolumns{10}
\tablecaption{Planets Younger than the Hyades}
\tablehead{
\colhead{KOI} & \colhead{\teff\ [K]} & \colhead{$EW_{\mathrm{Li}}$ [m\AA]} & \colhead{A(Li) [dex]} & \colhead{$\Delta_{\mathrm{A(Li),Hyades}}$ [dex]} & \colhead{Planet Number} & \colhead{$R_{\mathrm{p}}$ [$R_{\oplus}$]} & \colhead{Period [days]} &  \colhead{$F_{\mathrm{p}}$ [$F_{\oplus}$]}}
\def\arraystretch{1.0}
\startdata
1 & 5819 & 85.5 $\pm$ 7.9 & 2.62 $\pm$ 0.08 & 0.08 & 1 & $14.3^{+1.4}_{-1.4}$ & $2.5^{+0.0}_{-0.0}$ & $890.7^{+184.9}_{-184.9}$\\
2 & 6449 & 82.9 $\pm$ 7.3 & 3.11 $\pm$ 0.07 & 0.59 & 1 & $13.4^{+2.0}_{-2.0}$ & $2.2^{+0.0}_{-0.0}$ & $3029.6^{+931.2}_{-931.2}$\\
46 & 5661 & 98.2 $\pm$ 5.5 & 2.57 $\pm$ 0.07 & 0.47 & 1 & $5.7^{+0.7}_{-0.7}$ & $3.5^{+0.0}_{-0.0}$ & $1030.6^{+279.4}_{-279.4}$\\
\nodata & \nodata & \nodata & \nodata & \nodata & 2 & $1.2^{+0.2}_{-0.2}$ & $6.0^{+0.0}_{-0.0}$ & $497.1^{+132.2}_{-132.2}$\\
63 & 5673 & 90.3 $\pm$ 1.6 & 2.41 $\pm$ 0.06 & 0.27 & 1 & $6.1^{+0.5}_{-0.5}$ & $9.4^{+0.0}_{-0.0}$ & $109.0^{+18.3}_{-18.3}$\\
98 & 6500 & 52.1 $\pm$ 4.0 & 2.87 $\pm$ 0.06 & 0.62 & 1 & $8.5^{+1.0}_{-1.0}$ & $6.8^{+0.0}_{-0.0}$ & $572.2^{+142.3}_{-142.3}$\\
119 & 5681 & 100.6 $\pm$ 6.4 & 2.60 $\pm$ 0.07 & 0.43 & 2 & $7.2^{+0.9}_{-0.9}$ & $190.3^{+0.0}_{-0.0}$ & $7.1^{+1.9}_{-1.9}$\\
\nodata & \nodata & \nodata & \nodata & \nodata & 1 & $7.9^{+1.0}_{-1.0}$ & $49.2^{+0.0}_{-0.0}$ & $42.5^{+11.5}_{-11.5}$\\
149 & 5708 & 55.3 $\pm$ 7.2 & 2.27 $\pm$ 0.09 & 0.02 & 1 & $5.5^{+0.8}_{-0.8}$ & $14.6^{+0.0}_{-0.0}$ & $204.7^{+64.9}_{-64.9}$\\
323 & 5529 & 52.9 $\pm$ 5.4 & 1.98 $\pm$ 0.09 & 0.42 & 1 & $2.0^{+0.2}_{-0.2}$ & $5.8^{+0.0}_{-0.0}$ & $167.5^{+28.2}_{-28.2}$\\
331 & 5506 & 40.2 $\pm$ 6.8 & 1.91 $\pm$ 0.11 & 0.44 & 1 & $4.2^{+0.7}_{-0.7}$ & $18.7^{+0.0}_{-0.0}$ & $141.5^{+40.4}_{-40.4}$\\
660 & 5320 & 25.6 $\pm$ 5.4 & 1.49 $\pm$ 0.13 & 0.61 & 1 & $3.9^{+0.7}_{-0.7}$ & $6.1^{+0.0}_{-0.0}$ & $664.0^{+228.0}_{-228.0}$\\
684 & 5287 & 26.3 $\pm$ 5.0 & 1.47 $\pm$ 0.12 & 0.67 & 1 & $8.7^{+3.7}_{-3.7}$ & $4.0^{+0.0}_{-0.0}$ & $1201.6^{+389.0}_{-389.0}$\\
720 & 5260 & 14.5 $\pm$ 5.5 & 1.11 $\pm$ 0.23 & 0.39 & 1 & $3.0^{+0.2}_{-0.2}$ & $5.7^{+0.0}_{-0.0}$ & $130.6^{+22.2}_{-22.2}$\\
\nodata & \nodata & \nodata & \nodata & \nodata & 2 & $2.8^{+0.2}_{-0.2}$ & $10.0^{+0.0}_{-0.0}$ & $61.2^{+10.5}_{-10.5}$\\
\nodata & \nodata & \nodata & \nodata & \nodata & 4 & $1.6^{+0.1}_{-0.1}$ & $2.8^{+0.0}_{-0.0}$ & $337.1^{+57.2}_{-57.2}$\\
\nodata & \nodata & \nodata & \nodata & \nodata & 3 & $2.7^{+0.2}_{-0.2}$ & $18.4^{+0.0}_{-0.0}$ & $27.3^{+4.6}_{-4.6}$\\
853 & 4876 & 43.8 $\pm$ 11.0 & 1.17 $\pm$ 0.20 & 1.18 & 1 & $2.7^{+0.3}_{-0.3}$ & $8.2^{+0.0}_{-0.0}$ & $51.8^{+8.9}_{-8.9}$\\
\nodata & \nodata & \nodata & \nodata & \nodata & 2 & $2.5^{+0.4}_{-0.4}$ & $14.5^{+0.0}_{-0.0}$ & $24.2^{+4.1}_{-4.1}$\\
1117 & 6513 & 38.3 $\pm$ 6.4 & 2.72 $\pm$ 0.10 & 0.53 & 1 & $1.9^{+0.3}_{-0.3}$ & $11.1^{+0.0}_{-0.0}$ & $328.7^{+88.0}_{-88.0}$\\
1175 & 5640 & 46.8 $\pm$ 7.1 & 2.12 $\pm$ 0.10 & 0.08 & 1 & $2.6^{+0.4}_{-0.4}$ & $31.6^{+0.0}_{-0.0}$ & $89.8^{+25.7}_{-25.7}$\\
\nodata & \nodata & \nodata & \nodata & \nodata & 2 & $2.3^{+0.5}_{-0.5}$ & $17.2^{+0.0}_{-0.0}$ & $204.7^{+58.4}_{-58.4}$\\
1199 & 4772 & 85.9 $\pm$ 8.7 & 1.28 $\pm$ 0.13 & 1.40 & 1 & $2.5^{+0.2}_{-0.2}$ & $53.5^{+0.0}_{-0.0}$ & $4.0^{+0.7}_{-0.7}$\\
1208 & 6417 & 48.3 $\pm$ 6.1 & 2.77 $\pm$ 0.08 & 0.12 & 1 & $8.4^{+1.2}_{-1.2}$ & $700.0^{+0.0}_{-0.0}$ & $1.2^{+0.3}_{-0.3}$\\
1221 & 5002 & 9.0 $\pm$ 2.4 & 0.52 $\pm$ 0.16 & 0.36 & 1 & $4.7^{+0.7}_{-0.7}$ & $30.2^{+0.0}_{-0.0}$ & $133.6^{+39.3}_{-39.3}$\\
\nodata & \nodata & \nodata & \nodata & \nodata & 2 & $3.8^{+0.6}_{-0.6}$ & $51.1^{+0.0}_{-0.0}$ & $66.2^{+19.4}_{-19.4}$\\
1230$^a$ & 5119 & 26.8 $\pm$ 5.4 & 1.31 $\pm$ 0.13 & 0.92 & 1 & $38.7^{+6.6}_{-6.6}$ & $165.7^{+0.0}_{-0.0}$ & $30.4^{+10.6}_{-10.6}$\\
1413 & 5253 & 18.1 $\pm$ 8.3 & 1.26 $\pm$ 0.29 & 0.56 & 1 & $3.6^{+0.7}_{-0.7}$ & $12.6^{+0.0}_{-0.0}$ & $311.4^{+113.2}_{-113.2}$\\
\nodata & \nodata & \nodata & \nodata & \nodata & 2 & $3.0^{+0.5}_{-0.5}$ & $21.5^{+0.0}_{-0.0}$ & $153.6^{+55.8}_{-55.8}$\\
1438 & 5722 & 64.2 $\pm$ 9.5 & 2.36 $\pm$ 0.10 & 0.07 & 1 & $2.5^{+0.4}_{-0.4}$ & $6.9^{+0.0}_{-0.0}$ & $532.5^{+137.9}_{-137.9}$\\
1463 & 6532 & 38.2 $\pm$ 2.0 & 2.73 $\pm$ 0.05 & 0.66 & 1 & $23.0^{+2.8}_{-2.8}$ & $1064.3^{+0.0}_{-0.0}$ & $0.7^{+0.2}_{-0.2}$\\
1800 & 5621 & 100.1 $\pm$ 6.4 & 2.49 $\pm$ 0.08 & 0.51 & 1 & $6.3^{+0.5}_{-0.5}$ & $7.8^{+0.0}_{-0.0}$ & $126.0^{+21.2}_{-21.2}$\\
1839 & 5517 & 99.4 $\pm$ 5.3 & 2.28 $\pm$ 0.08 & 0.77 & 1 & $2.3^{+0.2}_{-0.2}$ & $9.6^{+0.0}_{-0.0}$ & $88.5^{+14.9}_{-14.9}$\\
\nodata & \nodata & \nodata & \nodata & \nodata & 2 & $2.3^{+0.2}_{-0.2}$ & $80.4^{+0.0}_{-0.0}$ & $5.2^{+0.9}_{-0.9}$\\
1864 & 5620 & 52.1 $\pm$ 7.7 & 2.15 $\pm$ 0.10 & 0.18 & 1 & $2.3^{+0.3}_{-0.3}$ & $3.2^{+0.0}_{-0.0}$ & $1482.2^{+431.3}_{-431.3}$\\
1985 & 4950 & 10.3 $\pm$ 5.9 & 0.62 $\pm$ 0.40 & 0.55 & 1 & $2.5^{+0.3}_{-0.3}$ & $5.8^{+0.0}_{-0.0}$ & $93.0^{+15.8}_{-15.8}$\\
2033 & 5051 & 12.9 $\pm$ 7.5 & 0.61 $\pm$ 0.40 & 0.35 & 1 & $1.5^{+0.1}_{-0.1}$ & $16.5^{+0.0}_{-0.0}$ & $25.3^{+4.3}_{-4.3}$\\
2035 & 5558 & 56.2 $\pm$ 5.5 & 2.06 $\pm$ 0.09 & 0.38 & 1 & $2.2^{+0.2}_{-0.2}$ & $1.9^{+0.0}_{-0.0}$ & $766.9^{+131.4}_{-131.4}$\\
2046 & 5579 & 44.4 $\pm$ 6.4 & 2.04 $\pm$ 0.10 & 0.25 & 1 & $2.5^{+0.4}_{-0.4}$ & $23.9^{+0.0}_{-0.0}$ & $70.7^{+22.3}_{-22.3}$\\
2115 & 5239 & 95.3 $\pm$ 23.1 & 1.90 $\pm$ 0.20 & 1.24 & 1 & $3.1^{+0.3}_{-0.3}$ & $15.7^{+0.0}_{-0.0}$ & $33.1^{+5.7}_{-5.7}$\\
2175 & 5459 & 43.8 $\pm$ 7.1 & 1.91 $\pm$ 0.11 & 0.61 & 1 & $3.0^{+0.4}_{-0.4}$ & $26.8^{+0.0}_{-0.0}$ & $84.7^{+25.3}_{-25.3}$\\
\nodata & \nodata & \nodata & \nodata & \nodata & 2 & $3.5^{+0.5}_{-0.5}$ & $72.4^{+0.0}_{-0.0}$ & $22.7^{+6.7}_{-6.7}$\\
2261 & 5176 & 31.4 $\pm$ 7.7 & 1.34 $\pm$ 0.17 & 0.83 & 1 & $1.2^{+0.1}_{-0.1}$ & $4.0^{+0.0}_{-0.0}$ & $193.2^{+33.0}_{-33.0}$\\
\nodata & \nodata & \nodata & \nodata & \nodata & 2 & $0.9^{+0.2}_{-0.2}$ & $6.6^{+0.0}_{-0.0}$ & $97.9^{+16.7}_{-16.7}$\\
2479 & 5372 & 28.7 $\pm$ 6.7 & 1.61 $\pm$ 0.14 & 0.59 & 1 & $2.2^{+0.4}_{-0.4}$ & $25.5^{+0.0}_{-0.0}$ & $93.5^{+33.1}_{-33.1}$\\
2516 & 5431 & 19.1 $\pm$ 8.1 & 1.48 $\pm$ 0.26 & 0.27 & 1 & $1.3^{+0.2}_{-0.2}$ & $2.8^{+0.0}_{-0.0}$ & $1611.6^{+440.9}_{-440.9}$\\
2541 & 5090 & 15.9 $\pm$ 6.9 & 0.96 $\pm$ 0.27 & 0.62 & 1 & $2.4^{+0.5}_{-0.5}$ & $7.4^{+0.0}_{-0.0}$ & $748.1^{+245.6}_{-245.6}$\\
\nodata & \nodata & \nodata & \nodata & \nodata & 2 & $2.6^{+0.4}_{-0.4}$ & $20.5^{+0.0}_{-0.0}$ & $192.6^{+63.3}_{-63.3}$\\
2639 & 5583 & 99.4 $\pm$ 13.2 & 2.50 $\pm$ 0.11 & 0.69 & 1 & $3.6^{+0.7}_{-0.7}$ & $25.1^{+0.0}_{-0.0}$ & $115.2^{+40.2}_{-40.2}$\\
\nodata & \nodata & \nodata & \nodata & \nodata & 2 & $9.7^{+61.1}_{-61.1}$ & $2.1^{+0.0}_{-0.0}$ & $3092.3^{+1075.3}_{-1075.3}$\\
2675 & 5756 & 82.1 $\pm$ 6.2 & 2.53 $\pm$ 0.07 & 0.13 & 1 & $2.2^{+0.2}_{-0.2}$ & $5.4^{+0.0}_{-0.0}$ & $243.0^{+40.8}_{-40.8}$\\
\nodata & \nodata & \nodata & \nodata & \nodata & 2 & $1.0^{+0.1}_{-0.1}$ & $1.1^{+0.0}_{-0.0}$ & $2017.0^{+341.1}_{-341.1}$\\
2678 & 5416 & 136.4 $\pm$ 6.4 & 2.42 $\pm$ 0.09 & 1.26 & 1 & $1.8^{+0.2}_{-0.2}$ & $3.8^{+0.0}_{-0.0}$ & $260.9^{+44.5}_{-44.5}$\\
2748 & 5499 & 33.0 $\pm$ 6.2 & 1.81 $\pm$ 0.12 & 0.37 & 1 & $2.5^{+0.4}_{-0.4}$ & $23.2^{+0.0}_{-0.0}$ & $95.0^{+27.1}_{-27.1}$\\
\nodata & \nodata & \nodata & \nodata & \nodata & 2 & $2.0^{+0.3}_{-0.3}$ & $5.8^{+0.0}_{-0.0}$ & $604.6^{+175.2}_{-175.2}$\\
2859 & 5260 & 12.4 $\pm$ 6.1 & 0.98 $\pm$ 0.32 & 0.26 & 1 & $0.7^{+0.1}_{-0.1}$ & $3.4^{+0.0}_{-0.0}$ & $272.5^{+46.3}_{-46.3}$\\
\nodata & \nodata & \nodata & \nodata & \nodata & 3 & $0.7^{+0.1}_{-0.1}$ & $4.3^{+0.0}_{-0.0}$ & $202.9^{+34.3}_{-34.3}$\\
\nodata & \nodata & \nodata & \nodata & \nodata & 4 & $0.7^{+0.1}_{-0.1}$ & $2.9^{+0.0}_{-0.0}$ & $342.1^{+58.6}_{-58.6}$\\
\nodata & \nodata & \nodata & \nodata & \nodata & 2 & $0.6^{+0.1}_{-0.1}$ & $2.0^{+0.0}_{-0.0}$ & $557.7^{+95.1}_{-95.1}$\\
3371 & 5428 & 54.1 $\pm$ 7.2 & 1.98 $\pm$ 0.10 & 0.78 & 1 & $1.3^{+0.1}_{-0.1}$ & $58.1^{+0.0}_{-0.0}$ & $7.4^{+1.3}_{-1.3}$\\
\nodata & \nodata & \nodata & \nodata & \nodata & 2 & $1.1^{+0.1}_{-0.1}$ & $12.3^{+0.0}_{-0.0}$ & $59.2^{+10.1}_{-10.1}$\\
3473$^b$ & 5157 & 82.6 $\pm$ 7.4 & 1.85 $\pm$ 0.11 & 1.38 & 1 & $7.0^{+2.8}_{-2.8}$ & $27.6^{+0.0}_{-0.0}$ & $14.4^{+2.5}_{-2.5}$\\
3835 & 5013 & 118.8 $\pm$ 6.1 & 1.92 $\pm$ 0.11 & 1.73 & 1 & $2.6^{+0.2}_{-0.2}$ & $47.1^{+0.0}_{-0.0}$ & $6.0^{+1.0}_{-1.0}$\\
3876 & 5720 & 119.8 $\pm$ 4.3 & 2.73 $\pm$ 0.06 & 0.45 & 1 & $1.9^{+0.2}_{-0.2}$ & $19.6^{+0.0}_{-0.0}$ & $42.4^{+7.2}_{-7.2}$\\
3886$^a$ & 4760 & 11.5 $\pm$ 6.3 & 0.45 $\pm$ 0.36 & 0.58 & 1 & $28.6^{+8.5}_{-8.5}$ & $5.6^{+0.0}_{-0.0}$ & $7866.2^{+4770.1}_{-4770.1}$\\
3891$^b$ & 5080 & 21.6 $\pm$ 5.2 & 1.08 $\pm$ 0.15 & 0.76 & 1 & $7.1^{+2.2}_{-2.2}$ & $47.1^{+0.0}_{-0.0}$ & $58.2^{+20.0}_{-20.0}$\\
3908 & 5721 & 58.1 $\pm$ 6.6 & 2.30 $\pm$ 0.09 & 0.02 & 1 & $2.9^{+0.4}_{-0.4}$ & $59.4^{+0.0}_{-0.0}$ & $37.4^{+10.8}_{-10.8}$\\
3936 & 5081 & 162.5 $\pm$ 4.8 & 2.03 $\pm$ 0.12 & 1.71 & 2 & $1.6^{+0.2}_{-0.2}$ & $13.0^{+0.0}_{-0.0}$ & $36.8^{+6.2}_{-6.2}$\\
3991 & 5606 & 96.7 $\pm$ 6.6 & 2.48 $\pm$ 0.08 & 0.56 & 1 & $1.4^{+0.2}_{-0.2}$ & $1.6^{+0.0}_{-0.0}$ & $1053.8^{+179.7}_{-179.7}$\\
4004 & 5739 & 78.6 $\pm$ 5.2 & 2.48 $\pm$ 0.07 & 0.13 & 1 & $1.1^{+0.1}_{-0.1}$ & $4.9^{+0.0}_{-0.0}$ & $256.4^{+43.6}_{-43.6}$\\
4146 & 5092 & 22.4 $\pm$ 5.9 & 0.67 $\pm$ 0.19 & 0.32 & 1 & $0.8^{+0.1}_{-0.1}$ & $3.5^{+0.0}_{-0.0}$ & $219.8^{+37.4}_{-37.4}$\\
4156 & 5807 & 96.8 $\pm$ 7.1 & 2.69 $\pm$ 0.07 & 0.18 & 1 & $1.4^{+0.2}_{-0.2}$ & $4.9^{+0.0}_{-0.0}$ & $786.5^{+209.0}_{-209.0}$\\
4226 & 5844 & 82.6 $\pm$ 7.6 & 2.62 $\pm$ 0.08 & 0.03 & 1 & $2.4^{+0.3}_{-0.3}$ & $49.6^{+0.0}_{-0.0}$ & $51.0^{+11.8}_{-11.8}$\\
4613 & 5443 & 23.6 $\pm$ 6.7 & 1.45 $\pm$ 0.17 & 0.20 & 1 & $0.7^{+0.1}_{-0.1}$ & $2.0^{+0.0}_{-0.0}$ & $674.4^{+115.6}_{-115.6}$\\
4647 & 5166 & 23.2 $\pm$ 9.6 & 1.18 $\pm$ 0.25 & 0.69 & 2 & $2.1^{+0.3}_{-0.3}$ & $12.0^{+0.0}_{-0.0}$ & $290.1^{+87.2}_{-87.2}$\\
\nodata & \nodata & \nodata & \nodata & \nodata & 1 & $2.5^{+0.4}_{-0.4}$ & $37.9^{+0.0}_{-0.0}$ & $62.1^{+18.7}_{-18.7}$\\
4663 & 5545 & 55.1 $\pm$ 6.6 & 2.12 $\pm$ 0.09 & 0.49 & 1 & $1.0^{+0.1}_{-0.1}$ & $5.7^{+0.0}_{-0.0}$ & $679.8^{+166.5}_{-166.5}$\\
4686 & 5698 & 75.6 $\pm$ 5.9 & 2.44 $\pm$ 0.07 & 0.22 & 1 & $1.2^{+0.2}_{-0.2}$ & $11.8^{+0.0}_{-0.0}$ & $262.5^{+60.0}_{-60.0}$\\
4745 & 4781 & 78.8 $\pm$ 14.6 & 1.41 $\pm$ 0.16 & 1.53 & 1 & $2.2^{+0.2}_{-0.2}$ & $177.7^{+0.0}_{-0.0}$ & $0.8^{+0.1}_{-0.1}$\\
4763 & 5695 & 52.4 $\pm$ 7.8 & 2.23 $\pm$ 0.10 & 0.02 & 1 & $2.0^{+0.3}_{-0.3}$ & $56.4^{+0.0}_{-0.0}$ & $30.2^{+8.7}_{-8.7}$\\
4775 & 5210 & 36.6 $\pm$ 6.3 & 1.48 $\pm$ 0.11 & 0.90 & 1 & $1.9^{+0.3}_{-0.3}$ & $16.4^{+0.0}_{-0.0}$ & $177.1^{+56.9}_{-56.9}$\\
4811 & 5572 & 39.9 $\pm$ 9.4 & 1.97 $\pm$ 0.15 & 0.22 & 1 & $2.5^{+0.4}_{-0.4}$ & $21.7^{+0.0}_{-0.0}$ & $121.5^{+37.9}_{-37.9}$\\
4834 & 5030 & 19.5 $\pm$ 5.4 & 0.86 $\pm$ 0.16 & 0.64 & 1 & $1.4^{+0.2}_{-0.2}$ & $3.7^{+0.0}_{-0.0}$ & $1420.4^{+427.7}_{-427.7}$\\
5057 & 5004 & 26.5 $\pm$ 7.1 & 1.17 $\pm$ 0.16 & 1.01 & 1 & $11.0^{+2.4}_{-2.4}$ & $493.7^{+0.0}_{-0.0}$ & $7.4^{+3.2}_{-3.2}$\\
5107 & 4933 & 27.4 $\pm$ 10.4 & 1.11 $\pm$ 0.24 & 1.05 & 1 & $18.1^{+3.2}_{-3.2}$ & $169.3^{+0.0}_{-0.0}$ & $71.1^{+23.8}_{-23.8}$\\
5119 & 4984 & 13.7 $\pm$ 7.3 & 0.84 $\pm$ 0.35 & 0.71 & 1 & $12.7^{+3.0}_{-3.0}$ & $143.2^{+0.0}_{-0.0}$ & $35.1^{+13.2}_{-13.2}$\\
\enddata
\tablecomments{Planets in systems with $\Delta_{\mathrm{A(Li),Hyades}} > 0.0$, where $\Delta_{\mathrm{A(Li),Hyades}}$ is the A(Li) of each point subtracted by the Hyades curve at each \teff (systems younger than the Hyades). Upper limits in A(Li) are excluded. We list stellar data, produced from our pipeline, in the first five columns, while we list the individual planet information from the CKS in the last four columns. Errors in \teff\ are 60 K. The last four columns, in order, include the KOI planet number (which is appended to the KOI number), planet radius in Earth radii, planet orbital period in days, and planet insolation flux in Earth flux. All planets in this table are either confirmed planets (P$_{\mathrm{planet}} \gtrsim 0.99$) or planet candidates (P$_{\mathrm{planet}} \gtrsim 0.90$) -- no false positives are included. The errors in the planet orbital periods from the Archive are orders of magnitude smaller than the listed precision. Any \nodata marks indicate an additional planet which belongs to the system listed above.}
\tablenotetext{a}{These planet sizes are too large to be physical. After further investigation, we found that KOI 1230's companion has been dispositioned as a certified false positive since we last accessed the NASA Exoplanet Archive. KOI 3886's companion has also been dispositioned as a certified false positive.}
\tablenotetext{b}{These systems have been modified with manually calculated $R_{\mathrm{p}}$ based on the transit depth and the stellar radius due to a non-physical planet radius in the CKS catalog.} \label{tab:young}
\end{deluxetable*}

\bibliography{ApJLiCKS}

\begin{thebibliography}{}
\expandafter\ifx\csname natexlab\endcsname\relax\def\natexlab#1{#1}\fi

\bibitem[{{Andersen} {et~al.}(1984){Andersen}, {Gustafsson}, \&
  {Lambert}}]{Andersen1984}
{Andersen}, J., {Gustafsson}, B., \& {Lambert}, D.~L. 1984, \aap, 136, 65

\bibitem[{{Asplund} {et~al.}(2009){Asplund}, {Grevesse}, {Sauval}, \&
  {Scott}}]{Asplund2009}
{Asplund}, M., {Grevesse}, N., {Sauval}, A.~J., \& {Scott}, P. 2009, \araa, 47,
  481

\bibitem[{{Babu} \& {Feigelson}(2006)}]{Babu2006}
{Babu}, G.~J., \& {Feigelson}, E.~D. 2006, in Astronomical Society of the
  Pacific Conference Series, Vol. 351, Astronomical Data Analysis Software and
  Systems XV, ed. C.~{Gabriel}, C.~{Arviset}, D.~{Ponz}, \& S.~{Enrique}, 127

\bibitem[{{Baumann} {et~al.}(2010){Baumann}, {Ram{\'{\i}}rez}, {Mel{\'e}ndez},
  {Asplund}, \& {Lind}}]{Baumann2010}
{Baumann}, P., {Ram{\'{\i}}rez}, I., {Mel{\'e}ndez}, J., {Asplund}, M., \&
  {Lind}, K. 2010, \aap, 519, A87

\bibitem[{{Bertran de Lis} {et~al.}(2015){Bertran de Lis}, {Delgado Mena},
  {Adibekyan}, {Santos}, \& {Sousa}}]{Lis2015}
{Bertran de Lis}, S., {Delgado Mena}, E., {Adibekyan}, V.~Z., {Santos}, N.~C.,
  \& {Sousa}, S.~G. 2015, \aap, 576, A89

\bibitem[{{Boesgaard} {et~al.}(2016){Boesgaard}, {Lum}, {Deliyannis}, {King},
  {Pinsonneault}, \& {Somers}}]{Boesgaard2016}
{Boesgaard}, A.~M., {Lum}, M.~G., {Deliyannis}, C.~P., {et~al.} 2016, \apj,
  830, 49

\bibitem[{{Borucki} {et~al.}(2013){Borucki}, {Agol}, {Fressin}, {Kaltenegger},
  {Rowe}, {Isaacson}, {Fischer}, {Batalha}, {Lissauer}, {Marcy}, {Fabrycky},
  {D{\'e}sert}, {Bryson}, {Barclay}, {Bastien}, {Boss}, {Brugamyer},
  {Buchhave}, {Burke}, {Caldwell}, {Carter}, {Charbonneau}, {Crepp},
  {Christensen-Dalsgaard}, {Christiansen}, {Ciardi}, {Cochran}, {DeVore},
  {Doyle}, {Dupree}, {Endl}, {Everett}, {Ford}, {Fortney}, {Gautier}, {Geary},
  {Gould}, {Haas}, {Henze}, {Howard}, {Howell}, {Huber}, {Jenkins}, {Kjeldsen},
  {Kolbl}, {Kolodziejczak}, {Latham}, {Lee}, {Lopez}, {Mullally}, {Orosz},
  {Prsa}, {Quintana}, {Sanchis-Ojeda}, {Sasselov}, {Seader}, {Shporer},
  {Steffen}, {Still}, {Tenenbaum}, {Thompson}, {Torres}, {Twicken}, {Welsh}, \&
  {Winn}}]{Borucki2013}
{Borucki}, W.~J., {Agol}, E., {Fressin}, F., {et~al.} 2013, Science, 340, 587

\bibitem[{{Brandt} \& {Huang}(2015)}]{Brandt2015}
{Brandt}, T.~D., \& {Huang}, C.~X. 2015, \apj, 807, 24

\bibitem[{{Cayrel}(1988)}]{Cayrel1988}
{Cayrel}, R. 1988, in IAU Symposium, Vol. 132, The Impact of Very High S/N
  Spectroscopy on Stellar Physics, ed. G.~{Cayrel de Strobel} \& M.~{Spite},
  345

\bibitem[{{Chaussidon}(2007)}]{Chaussidon2007}
{Chaussidon}, M. 2007, {Formation of the Solar System: a Chronology Based on
  Meteorites}, ed. M.~{Gargaud}, H.~{Martin}, \& P.~{Claeys}, 45

\bibitem[{{Coughlin} {et~al.}(2016){Coughlin}, {Mullally}, {Thompson}, {Rowe},
  {Burke}, {Latham}, {Batalha}, {Ofir}, {Quarles}, {Henze}, {Wolfgang},
  {Caldwell}, {Bryson}, {Shporer}, {Catanzarite}, {Akeson}, {Barclay},
  {Borucki}, {Boyajian}, {Campbell}, {Christiansen}, {Girouard}, {Haas},
  {Howell}, {Huber}, {Jenkins}, {Li}, {Patil-Sabale}, {Quintana}, {Ramirez},
  {Seader}, {Smith}, {Tenenbaum}, {Twicken}, \& {Zamudio}}]{Coughlin2016}
{Coughlin}, J.~L., {Mullally}, F., {Thompson}, S.~E., {et~al.} 2016, \apjs,
  224, 12

\bibitem[{{Cumming} {et~al.}(2008){Cumming}, {Butler}, {Marcy}, {Vogt},
  {Wright}, \& {Fischer}}]{Cumming2008}
{Cumming}, A., {Butler}, R.~P., {Marcy}, G.~W., {et~al.} 2008, \pasp, 120, 531

\bibitem[{{Delgado Mena} {et~al.}(2014){Delgado Mena}, {Israelian},
  {Gonz{\'a}lez Hern{\'a}ndez}, {Sousa}, {Mortier}, {Santos}, {Adibekyan},
  {Fernandes}, {Rebolo}, {Udry}, \& {Mayor}}]{Mena2014}
{Delgado Mena}, E., {Israelian}, G., {Gonz{\'a}lez Hern{\'a}ndez}, J.~I.,
  {et~al.} 2014, \aap, 562, A92

\bibitem[{{Delgado Mena} {et~al.}(2015){Delgado Mena}, {Bertr{\'a}n de Lis},
  {Adibekyan}, {Sousa}, {Figueira}, {Mortier}, {Gonz{\'a}lez Hern{\'a}ndez},
  {Tsantaki}, {Israelian}, \& {Santos}}]{Mena2015}
{Delgado Mena}, E., {Bertr{\'a}n de Lis}, S., {Adibekyan}, V.~Z., {et~al.}
  2015, \aap, 576, A69

\bibitem[{Engmann \& Cousineau(2011)}]{Engmann2011}
Engmann, S., \& Cousineau, D. 2011, Journal of Applied Quantitative Methods, 6,
  1

\bibitem[{{Figueira} {et~al.}(2014){Figueira}, {Faria}, {Delgado-Mena},
  {Adibekyan}, {Sousa}, {Santos}, \& {Israelian}}]{Figueira2014}
{Figueira}, P., {Faria}, J.~P., {Delgado-Mena}, E., {et~al.} 2014, \aap, 570,
  A21

\bibitem[{{Fortney} {et~al.}(2007){Fortney}, {Marley}, \&
  {Barnes}}]{Fortney2007}
{Fortney}, J.~J., {Marley}, M.~S., \& {Barnes}, J.~W. 2007, \apj, 659, 1661

\bibitem[{{Fressin} {et~al.}(2013){Fressin}, {Torres}, {Charbonneau}, {Bryson},
  {Christiansen}, {Dressing}, {Jenkins}, {Walkowicz}, \&
  {Batalha}}]{Fressin2013}
{Fressin}, F., {Torres}, G., {Charbonneau}, D., {et~al.} 2013, \apj, 766, 81

\bibitem[{{Fulton} {et~al.}(2017){Fulton}, {Petigura}, {Howard}, {Isaacson},
  {Marcy}, {Cargile}, {Hebb}, {Weiss}, {Johnson}, {Morton}, {Sinukoff},
  {Crossfield}, \& {Hirsch}}]{Fulton2017}
{Fulton}, B.~J., {Petigura}, E.~A., {Howard}, A.~W., {et~al.} 2017, ArXiv
  e-prints, arXiv:1703.10375

\bibitem[{{Gonzalez}(2014)}]{Gonzalez2014}
{Gonzalez}, G. 2014, \mnras, 441, 1201

\bibitem[{{Gonzalez}(2015)}]{Gonzalez2015}
---. 2015, \mnras, 446, 1020

\bibitem[{{Grunblatt} {et~al.}(2016){Grunblatt}, {Huber}, {Gaidos}, {Lopez},
  {Fulton}, {Vanderburg}, {Barclay}, {Fortney}, {Howard}, {Isaacson}, {Mann},
  {Petigura}, {Silva Aguirre}, \& {Sinukoff}}]{Grunblatt2016}
{Grunblatt}, S.~K., {Huber}, D., {Gaidos}, E.~J., {et~al.} 2016, \aj, 152, 185

\bibitem[{{Gustafsson} {et~al.}(2008){Gustafsson}, {Edvardsson}, {Eriksson},
  {J{\o}rgensen}, {Nordlund}, \& {Plez}}]{Gustafsson2008}
{Gustafsson}, B., {Edvardsson}, B., {Eriksson}, K., {et~al.} 2008, \aap, 486,
  951

\bibitem[{{Herbig}(1965)}]{Herbig1965}
{Herbig}, G.~H. 1965, \apj, 141, 588

\bibitem[{{Howard} {et~al.}(2010){Howard}, {Marcy}, {Johnson}, {Fischer},
  {Wright}, {Isaacson}, {Valenti}, {Anderson}, {Lin}, \& {Ida}}]{Howard2010}
{Howard}, A.~W., {Marcy}, G.~W., {Johnson}, J.~A., {et~al.} 2010, Science, 330,
  653

\bibitem[{{Howard} {et~al.}(2012){Howard}, {Marcy}, {Bryson}, {Jenkins},
  {Rowe}, {Batalha}, {Borucki}, {Koch}, {Dunham}, {Gautier}, {Van Cleve},
  {Cochran}, {Latham}, {Lissauer}, {Torres}, {Brown}, {Gilliland}, {Buchhave},
  {Caldwell}, {Christensen-Dalsgaard}, {Ciardi}, {Fressin}, {Haas}, {Howell},
  {Kjeldsen}, {Seager}, {Rogers}, {Sasselov}, {Steffen}, {Basri},
  {Charbonneau}, {Christiansen}, {Clarke}, {Dupree}, {Fabrycky}, {Fischer},
  {Ford}, {Fortney}, {Tarter}, {Girouard}, {Holman}, {Johnson}, {Klaus},
  {Machalek}, {Moorhead}, {Morehead}, {Ragozzine}, {Tenenbaum}, {Twicken},
  {Quinn}, {Isaacson}, {Shporer}, {Lucas}, {Walkowicz}, {Welsh}, {Boss},
  {Devore}, {Gould}, {Smith}, {Morris}, {Prsa}, {Morton}, {Still}, {Thompson},
  {Mullally}, {Endl}, \& {MacQueen}}]{Howard2012}
{Howard}, A.~W., {Marcy}, G.~W., {Bryson}, S.~T., {et~al.} 2012, \apjs, 201, 15

\bibitem[{{Huber} {et~al.}(2014){Huber}, {Silva Aguirre}, {Matthews},
  {Pinsonneault}, {Gaidos}, {Garc{\'{\i}}a}, {Hekker}, {Mathur}, {Mosser},
  {Torres}, {Bastien}, {Basu}, {Bedding}, {Chaplin}, {Demory}, {Fleming},
  {Guo}, {Mann}, {Rowe}, {Serenelli}, {Smith}, \& {Stello}}]{Huber2014}
{Huber}, D., {Silva Aguirre}, V., {Matthews}, J.~M., {et~al.} 2014, \apjs, 211,
  2

\bibitem[{Hunter(2007)}]{Hunter2007}
Hunter, J.~D. 2007, Computing In Science \& Engineering, 9, 90

\bibitem[{{Israelian} {et~al.}(2009){Israelian}, {Delgado Mena}, {Santos},
  {Sousa}, {Mayor}, {Udry}, {Dom{\'{\i}}nguez Cerde{\~n}a}, {Rebolo}, \&
  {Randich}}]{Israelian2009}
{Israelian}, G., {Delgado Mena}, E., {Santos}, N.~C., {et~al.} 2009, \nat, 462,
  189

\bibitem[{{Johnson} {et~al.}(2017){Johnson}, {Petigura}, {Fulton}, {Marcy},
  {Howard}, {Isaacson}, {Hebb}, {Cargile}, {Morton}, {Weiss}, {Winn}, {Rogers},
  {Sinukoff}, \& {Hirsch}}]{Johnson2017}
{Johnson}, J.~A., {Petigura}, E.~A., {Fulton}, B.~J., {et~al.} 2017, ArXiv
  e-prints, arXiv:1703.10402

\bibitem[{Jones {et~al.}(2001--)Jones, Oliphant, Peterson, {et~al.}}]{Scipy}
Jones, E., Oliphant, T., Peterson, P., {et~al.} 2001--, {SciPy}: Open source
  scientific tools for {Python}, , , [Online; accessed 2018-01-19]

\bibitem[{{Lodders} {et~al.}(2009){Lodders}, {Palme}, \& {Gail}}]{Lodders2009}
{Lodders}, K., {Palme}, H., \& {Gail}, H.-P. 2009, Landolt B{\"o}rnstein,
  arXiv:0901.1149

\bibitem[{{Lopez} \& {Fortney}(2013)}]{Lopez2013}
{Lopez}, E.~D., \& {Fortney}, J.~J. 2013, \apj, 776, 2

\bibitem[{{Lopez} {et~al.}(2012){Lopez}, {Fortney}, \& {Miller}}]{Lopez2012}
{Lopez}, E.~D., {Fortney}, J.~J., \& {Miller}, N. 2012, \apj, 761, 59

\bibitem[{{Mayor} {et~al.}(2011){Mayor}, {Marmier}, {Lovis}, {Udry},
  {S{\'e}gransan}, {Pepe}, {Benz}, {Bertaux}, {Bouchy}, {Dumusque}, {Lo Curto},
  {Mordasini}, {Queloz}, \& {Santos}}]{Mayor2011}
{Mayor}, M., {Marmier}, M., {Lovis}, C., {et~al.} 2011, ArXiv e-prints,
  arXiv:1109.2497

\bibitem[{McKinney(2010)}]{Pandas}
McKinney, W. 2010, in Proceedings of the 9th Python in Science Conference, ed.
  S.~van~der Walt \& J.~Millman, 51 -- 56

\bibitem[{Newville {et~al.}(2014)Newville, Stensitzki, Allen, \&
  Ingargiola}]{Newville2014}
Newville, M., Stensitzki, T., Allen, D.~B., \& Ingargiola, A. 2014, {LMFIT:
  Non-Linear Least-Square Minimization and Curve-Fitting for Python�}, , ,
  doi:10.5281/zenodo.11813

\bibitem[{{Orosz} {et~al.}(2012){Orosz}, {Welsh}, {Carter}, {Fabrycky},
  {Cochran}, {Endl}, {Ford}, {Haghighipour}, {MacQueen}, {Mazeh},
  {Sanchis-Ojeda}, {Short}, {Torres}, {Agol}, {Buchhave}, {Doyle}, {Isaacson},
  {Lissauer}, {Marcy}, {Shporer}, {Windmiller}, {Barclay}, {Boss}, {Clarke},
  {Fortney}, {Geary}, {Holman}, {Huber}, {Jenkins}, {Kinemuchi}, {Kruse},
  {Ragozzine}, {Sasselov}, {Still}, {Tenenbaum}, {Uddin}, {Winn}, {Koch}, \&
  {Borucki}}]{Orosz2012}
{Orosz}, J.~A., {Welsh}, W.~F., {Carter}, J.~A., {et~al.} 2012, Science, 337,
  1511

\bibitem[{{Perryman} {et~al.}(1998){Perryman}, {Brown}, {Lebreton}, {Gomez},
  {Turon}, {Cayrel de Strobel}, {Mermilliod}, {Robichon}, {Kovalevsky}, \&
  {Crifo}}]{Perryman1998}
{Perryman}, M.~A.~C., {Brown}, A.~G.~A., {Lebreton}, Y., {et~al.} 1998, \aap,
  331, 81

\bibitem[{{Petigura}(2015)}]{Petigura2015}
{Petigura}, E.~A. 2015, PhD thesis, University of California, Berkeley

\bibitem[{{Petigura} {et~al.}(2017){Petigura}, {Howard}, {Marcy}, {Johnson},
  {Isaacson}, {Cargile}, {Hebb}, {Fulton}, {Weiss}, {Morton}, {Winn}, {Rogers},
  {Sinukoff}, {Hirsch}, \& {Crossfield}}]{Petigura2017}
{Petigura}, E.~A., {Howard}, A.~W., {Marcy}, G.~W., {et~al.} 2017, ArXiv
  e-prints, arXiv:1703.10400

\bibitem[{{Ram{\'{\i}}rez} {et~al.}(2012){Ram{\'{\i}}rez}, {Fish}, {Lambert},
  \& {Allende Prieto}}]{Ramirez2012}
{Ram{\'{\i}}rez}, I., {Fish}, J.~R., {Lambert}, D.~L., \& {Allende Prieto}, C.
  2012, \apj, 756, 46

\bibitem[{{Reddy} {et~al.}(2002){Reddy}, {Lambert}, {Laws}, {Gonzalez}, \&
  {Covey}}]{Reddy2002}
{Reddy}, B.~E., {Lambert}, D.~L., {Laws}, C., {Gonzalez}, G., \& {Covey}, K.
  2002, \mnras, 335, 1005

\bibitem[{{Sanchis-Ojeda} {et~al.}(2013){Sanchis-Ojeda}, {Rappaport}, {Winn},
  {Levine}, {Kotson}, {Latham}, \& {Buchhave}}]{Sanchis-Ojeda2013}
{Sanchis-Ojeda}, R., {Rappaport}, S., {Winn}, J.~N., {et~al.} 2013, \apj, 774,
  54

\bibitem[{{Sansonetti} {et~al.}(1995){Sansonetti}, {Richou}, {Engleman}, \&
  {Radziemski}}]{Sansonetti1995}
{Sansonetti}, C.~J., {Richou}, B., {Engleman}, Jr., R., \& {Radziemski}, L.~J.
  1995, \pra, 52, 2682

\bibitem[{{Science Software Branch at STScI}(2012)}]{Pyraf2012}
{Science Software Branch at STScI}. 2012, {PyRAF: Python alternative for IRAF},
  Astrophysics Source Code Library, , , ascl:1207.011

\bibitem[{{Smith} {et~al.}(1998){Smith}, {Lambert}, \& {Nissen}}]{Smith1998}
{Smith}, V.~V., {Lambert}, D.~L., \& {Nissen}, P.~E. 1998, \apj, 506, 405

\bibitem[{{Sneden} {et~al.}(2012){Sneden}, {Bean}, {Ivans}, {Lucatello}, \&
  {Sobeck}}]{Sneden2012}
{Sneden}, C., {Bean}, J., {Ivans}, I., {Lucatello}, S., \& {Sobeck}, J. 2012,
  {MOOG: LTE line analysis and spectrum synthesis}, Astrophysics Source Code
  Library, , , ascl:1202.009

\bibitem[{{Soderblom}(2010)}]{Soderblom2010}
{Soderblom}, D.~R. 2010, \araa, 48, 581

\bibitem[{{Sousa} {et~al.}(2010){Sousa}, {Fernandes}, {Israelian}, \&
  {Santos}}]{Sousa2010}
{Sousa}, S.~G., {Fernandes}, J., {Israelian}, G., \& {Santos}, N.~C. 2010,
  \aap, 512, L5

\bibitem[{{Takeda} {et~al.}(2013){Takeda}, {Honda}, {Ohnishi}, {Ohkubo},
  {Hirata}, \& {Sadakane}}]{Takeda2013}
{Takeda}, Y., {Honda}, S., {Ohnishi}, T., {et~al.} 2013, \pasj, 65,
  arXiv:1212.6318

\bibitem[{{Valenti} \& {Piskunov}(2012)}]{Valenti2012}
{Valenti}, J.~A., \& {Piskunov}, N. 2012, {SME: Spectroscopy Made Easy},
  Astrophysics Source Code Library, , , ascl:1202.013

\bibitem[{{Wallace} {et~al.}(2011){Wallace}, {Hinkle}, {Livingston}, \&
  {Davis}}]{Hinkle2011}
{Wallace}, L., {Hinkle}, K.~H., {Livingston}, W.~C., \& {Davis}, S.~P. 2011,
  \apjs, 195, 6

\bibitem[{{Xiong} \& {Deng}(2009)}]{Xiong2009}
{Xiong}, D.~R., \& {Deng}, L. 2009, \mnras, 395, 2013

\bibitem[{{Yan} \& {Drake}(1995)}]{Yan1995}
{Yan}, Z.-C., \& {Drake}, G.~W.~F. 1995, \pra, 52, R4316

\end{thebibliography}

\end{document}